\documentclass[a4paper,11pt]{article}
\usepackage{jheppub} 
\usepackage{xcolor}
\usepackage{physics}
\usepackage{xspace}
\usepackage{mathrsfs}
\usepackage{soul}
\usepackage{mathtools}
\usepackage{tensor}
\usepackage{tabularx}
\usepackage{cancel}
\usepackage{booktabs}
\usepackage{ytableau}
\usepackage{marginnote}

\usepackage{amsthm}

\makeatletter
\g@addto@macro\bfseries{\boldmath}
\makeatother

\ytableausetup{boxsize=0.3em,centertableaux}

\newcommand{\two}{{\ydiagram{1,1}}}
\newcommand{\twotwo}{{\ydiagram{2,2}}}
\newcommand{\threeone}{{\ydiagram{2,1,1}}}
\newcommand{\four}{{\ydiagram{1,1,1,1}}}
\newcommand{\six}{{\ydiagram{1,1,1,1,1,1}}}
\newcommand{\fiveone}{{\ydiagram{2,1,1,1,1}}}
\newcommand{\fourtwo}{{\ydiagram{2,2,1,1}}}
\newcommand{\five}{{\ydiagram{1,1,1,1,1}}}
\newcommand{\threethree}{{\ydiagram{2,2,2}}}
\newcommand{\threetwoone}{{\ydiagram{3,2,1}}}
\newcommand{\fouroneone}{{\ydiagram{3,1,1,1}}}
\newcommand{\twotwotwo}{{\ydiagram{3,3}}}
\newcommand{\oneone}{{\ydiagram{2}}}
\newcommand{\twooneone}{{\ydiagram{3,1}}}
\newcommand{\fourone}{{\ydiagram{2,1,1,1}}}
\newcommand{\three}{{\ydiagram{1,1,1}}}
\newcommand{\one}{{\ydiagram{1}}}
\newcommand{\threetwo}{{\ydiagram{2,2,1}}}
\newcommand{\twoone}{{\ydiagram{2,1}}}

\newcommand{\A}{\mathcal{A}}
\newcommand{\B}{\mathcal{B}}
\newcommand{\C}{\mathcal{C}}
\newcommand{\D}{\mathcal{D}}
\newcommand{\M}{\mathcal{M}}
\newcommand{\pq}[1]{${\lfloor #1\rfloor}$}
\newcommand{\pqi}[1]{${[#1]}$}
\preprint{Imperial-TP-2024-CH-04\\ \rightline{UUITP-11/24}}
\title{Gauge-invariant magnetic charges in linearised gravity}

\author[a]{Chris Hull,}
\author[a]{Maxwell L. Hutt,}
\author[a,b]{Ulf Lindstr\"{o}m}

\affiliation[a]{The Blackett Laboratory, Imperial College London, Prince Consort Road, London, SW7 2AZ, UK}
\affiliation[b]{Department of Physics and Astronomy, Uppsala University,
Box 516, SE-75120 Uppsala, Sweden
and Centre for Geometry and Physics, Uppsala University,
Box 480, SE-75106 Uppsala, Sweden
}

\emailAdd{c.hull@imperial.ac.uk, m.hutt22@imperial.ac.uk, ulf.lindstrom@physics.uu.se}

\abstract{
Linearised gravity has magnetic charges carried by (linearised) Kaluza-Klein monopoles. A gauge-invariant expression is found for these charges that is similar to Penrose's gauge-invariant expression for the ADM charges. A systematic search is made for other gauge-invariant charges.
}

\begin{document}
\maketitle
\flushbottom

\section{Introduction}

For the linearised theory of the graviton in Minkowski space, each Killing vector of Minkowski space gives an invariance of the theory and there is a corresponding conserved charge $Q[k]$ that can be constructed via a linearisation of the ADM construction \cite{Abbott1982StabilityConstant}.
In Ref.~\cite{HullYetAppear} analogous `electric' conserved charges $Q[\kappa]$, $Q[\lambda]$ were constructed in spacetime dimensions $d\geq 5$ for the dual graviton theory \cite{Hull2000,Hull2001DualityFields} corresponding to certain generalised Killing tensors $\kappa$, $\lambda$ that give the invariances of the dual theory. These were then reinterpreted as `magnetic' charges for the graviton theory, for which they are topological charges.

In Ref.~\cite{Penrose1982Quasi-localRelativity}, covariant 2-form currents 
\begin{equation}\label{eq:Penrose2form}
    Y[K]_{\mu\nu} = R_{\mu\nu \alpha\beta} K^ {\alpha\beta}
\end{equation}
were introduced and shown to be conserved in regions without sources provided that $K_{\alpha\beta}$ is a conformal Killing-Yano (CKY) tensor. 
Integrating these over a closed codimension-2 surface then gives conserved charges, which we refer to as Penrose charges.
These were studied in Ref.~\cite{PAPER1} and shown to give a linear combination of $Q[k]$, $Q[\lambda]$ and a topological charge, with $k$ and $\lambda$ constructed from $K$. 
For the graviton theory, the Penrose charges can be interpreted as giving a covariantisation of $Q[k]$. However, the Penrose charges can also be interpreted in the dual graviton theory as a covariantisation of the charge $Q[\lambda]$. We will discuss the charges in the dual graviton theory in a forthcoming publication and will here focus on the charges in the graviton theory.

Kaluza-Klein monopoles play an important role in string theory \cite{Hull:Uduality} and carry a gravitational magnetic charge \cite{ Hull:braneCharges}.
The magnetic charge $Q[\kappa]$ in the linearised theory is the charge carried by linearised Kaluza-Klein monopoles \cite{HullYetAppear}.
The magnetic charge $Q[\kappa]$, however, is not related to the Penrose charges.
One of the aims of this paper is to find a covariant current that gives this charge.
This leads us to generalise the Penrose construction and seek conserved 2-form currents, linear in the Riemann tensor, of the form
\begin{equation}\label{eq:Omega[V]_intro}
    \Omega[V]_{\mu\nu} = R^{\alpha\beta\gamma\delta} V_{\mu\nu\alpha\beta|\gamma\delta}
\end{equation}
for some rank-6 tensor $V_{\mu\nu\alpha\beta|\gamma\delta}$ with the symmetries\footnote{We use a vertical bar to separate groups of fully antisymmetric indices.}
\begin{equation}
    V_{\mu\nu\alpha\beta|\gamma\delta} = V_{[\mu\nu\alpha\beta]|[\gamma\delta]}
\end{equation}
We investigate the conditions on $V$ for this current to be conserved, classifying the full set of such conserved 2-forms.
One of our main results is that, as well as the Penrose charges, the $\kappa$-type charges can also be constructed as a subset of these charges, and so also have a covariant expression.

In the quantum theory, gauge-invariant operators are particularly important. Finding covariant forms for electric and magnetic gravitational charges, which is our goal here, then gives gauge invariant topological operators that generate continuous generalised symmetries.

One motivation of this work is to further understand the generalised symmetries of the free graviton theory, following work by \cite{Hinterbichler2023GravitySymmetries}. There, it was shown that any system containing specific electric and magnetic `bi-form symmetries' which participate in a particular mixed 't Hooft anomaly will have a gapless phase with a spin-2 mode in the IR spectrum. In other words, the graviton can be viewed as the Nambu-Goldstone boson associated with the non-linear realisation of such symmetries. Similar ideas were used in \cite{Hinterbichler:2024cxn} to show that certain anomalies cannot be realised by any local, unitary theory. Finding gauge-invariant expressions for the conserved quantities related to all the (generalised) symmetries of the theory and understanding their anomalies is, therefore, a promising avenue to gain insight into the IR physics at play.
Other work on generalised symmetries in gravity can be found in \cite{Benedetti2022GeneralizedGraviton, Benedetti2023GeneralizedGravitons, BenedettiNoether, Gomez-Fayren2023CovariantRelativity, Cheung:2024ypq}.

Consider first the symmetries corresponding to the Killing vectors $k$. For each $k$ there is a 1-form current
\begin{equation}\label{eq:j[k]1}
     j[k]_\mu  \equiv T_{\mu\nu} k^\nu 
\end{equation}
where the energy-momentum tensor $ T_{\mu\nu}$ is the source for the linearised Einstein equation.
This can be integrated over a co-dimension one surface to give a conserved charge that generates a 0-form symmetry. However, there is a `secondary current' $J[k]_{\mu\nu}$ which satisfies
\begin{equation}\label{eq:j[k]2}
    \partial^\nu J[k]_{\mu\nu} =  j[k]_\mu
\end{equation}
so that this 2-form current is conserved wherever the source $j$ vanishes. However, the current $J[k]$ is not uniquely determined by \eqref{eq:j[k]2}. The ADM construction reviewed in~\cite{HullYetAppear} gives a non-covariant $J[k]$, but the Penrose construction gives a covariant current.

Integrating $J$ over a codimension 2 surface gives a conserved charge which in the quantum theory gives a topological operator -- it is unchanged under deformations of the surface that do not pass through regions in which $j\neq 0$. This charge generates a 1-form symmetry.
Note that whereas for non-linear gravity the surface needs to be taken at spatial or null infinity as gravitational energy-momentum is non-local, for the linearised theory it can be any surface as the energy-momentum is local.

There is a similar story for the $\kappa$ and $\lambda$ charges~\cite{HullYetAppear}.
In the dual theory, there is a primary 1-form current $j$ for each, and a corresponding 2-form current $J$ with $\partial^\nu J_{\mu\nu} =  j_\mu$, with corresponding charges and 0-form and 1-form symmetries, similarly to the case above.
The current $J$ gives rise to a 2-form current in the graviton theory, giving a topological charge there.
In this paper we will focus on the charges and postpone a discussion of the corresponding generalised symmetries to a future paper. 

Given the relevance of these charges to Kaluza-Klein monopoles, we also study the gauge-invariant charges constructed from $\Omega[V]$ on toroidally compactified spaces $\mathbb{R}^{1,D-1}\times T^n$ with $D+n=d$, giving them a unified higher-dimensional origin.

\subsection*{Conventions}

We refer to tensors whose indices can be split into fully anti-symmetric groups of lengths $n_1, n_2, \dots, n_r$ as \pq{n_1,n_2,\dots,n_r}-tensors. For example, a tensor with index symmetries $M_{\mu\nu\rho|\sigma|\alpha} = M_{[\mu\nu\rho]|\sigma|\alpha}$ is a \pq{3,1,1}-tensor. Such tensors have been studied in Refs.~\cite{Medeiros2003ExoticDuality, Bekaert2004, Dubois-Violette:1999iqe,Dubois-Violette:2000fok, Dubois-Violette:2001wjr,Howe2018SCKYT} where they are sometimes referred to as `multi-forms' and in the case of $r=2$ they are often called `bi-forms'.
Unless otherwise stated, these tensors are in \emph{reducible} $GL(d,\mathbb{R})$ representations. That is, they transform in a representation which is labelled by a tensor product of single-column Young tableaux of lengths $n_1,n_2,\dots,n_r$.

\emph{Irreducible} $GL(d,\mathbb{R})$ representations are instead labelled by \pqi{n_1,\dots,n_r} (note the different type of bracket).
The criterion for a reducible \pq{p,q}-tensor, $N$, to transform in an irreducible $GL(d,\mathbb{R})$ representation (i.e. one labelled by a single Young tableau with two columns of lengths $p$ and $q$) is
\begin{equation}\label{eq:irrep_constraint}
    N_{\mu_1\dots\mu_p|\nu_1\dots\nu_q} = N_{[\mu_1\dots\mu_p]|[\nu_1\dots\nu_q]} \qc N_{[\mu_1\dots\mu_p|\nu_1]\nu_2\dots\nu_q} = 0
\end{equation}
where we have assumed $p>q$. If $p=q$ then furthermore $N$ satisfies
\begin{equation}\label{eq:irrep_constraint2}
    N_{\mu_1\dots\mu_p|\nu_1\dots\nu_p} = N_{\nu_1\dots\nu_p|\mu_1\dots\mu_p}
\end{equation}

Irreducible representations of $GL(d,\mathbb{R})$ can be viewed as reducible representations of $SO(1,d-1)\subset GL(d,\mathbb{R})$ and can, therefore, be decomposed into a sum of irreducible representations of $SO(1,d-1)$. A \pqi{p,q}-tensor $A$ which transforms in an irreducible representation of $SO(1,d-1)$ labelled by a Young tableau with two columns of lengths $p$ and $q$ must satisfy eqs.~\eqref{eq:irrep_constraint} and \eqref{eq:irrep_constraint2}, and must also be traceless with respect to the $SO(1,d-1)$-invariant Minkowski metric $\eta_{\mu\nu}$:
\begin{equation}
    \eta^{\mu_1\nu_1}A_{\mu_1\dots\mu_p|\nu_1\dots\nu_q} = 0
\end{equation}

The Young symmetriser $\mathcal{Y}$ is a linear operator which projects a tensor onto a given irreducible representation. We will denote in the superscript whether the projection is to an irreducible $GL(d,\mathbb{R})$ or $SO(1,d-1)$ representation. We will denote in the subscript the Young tableau of the irreducible representation which is being projected on to. For example, $\mathcal{Y}_{\twoone}^{SO}$ is the projector onto the irreducible \pqi{2,1} representation of $SO(1,d-1)$ (so the resulting tensor is traceless).

We will use some of the 
operations on \pq{p,q}-tensors introduced in \cite{Medeiros2003ExoticDuality}.\footnote{A similar formalism was introduced in \cite{Bekaert2004}.}
The left and right exterior derivatives of a reducible \pq{p,q}-tensor $N$ are
\begin{align}
    (\dd_L A)_{\mu_1\dots\mu_{p+1}|\nu_1\dots\nu_q} &= \partial_{[\mu_1} A_{\mu_2\dots\mu_{p+1}]|\nu_1\dots\nu_q} \\
    (\dd_R A)_{\mu_1\dots\mu_p|\nu_1\dots\nu_{q+1}} &= A_{\mu_1\dots\mu_p|[\nu_1\dots\nu_q,\nu_{q+1}]}
\end{align}
where a comma denotes a partial derivative. These are nilpotent, $\dd_L^2 = \dd_R^2 = 0$, and commute, $[d_L,d_R]=0$. $N$ is $\dd_L$-closed if $\dd_L N = 0$ and $\dd_L$-exact if $N = \dd_L M$ for some $M$.
The $\dd_L$-cohomology of \pq{p,q}-tensors on a flat space is the set of $\dd_L$-closed \pq{p,q}-tensors modulo those which are $\dd_L$-exact.
We define $\dd_R$-closed and $\dd_R$-exact tensors, and so $\dd_R$-cohomology, analogously.

We also define left and right duality operations $\star_L$ and $\star_R$ by
\begin{align}
    (\star_L A)_{\mu_1\dots\mu_{d-p}|\nu_1\dots\nu_q} &= \frac{1}{p!} \epsilon_{\mu_1\dots\mu_{d-p}\alpha_1\dots\alpha_p} A\indices{^{\alpha_1\dots\alpha_p}_{|\nu_1\dots\nu_q}} \\
    (\star_R A)_{\mu_1\dots\mu_p|\nu_1\dots\nu_{d-q}} &= \frac{1}{q!} \epsilon_{\nu_1\dots\nu_{d-q}\alpha_1\dots\alpha_q} A\indices{_{\mu_1\dots\mu_p|}^{\alpha_1\dots\alpha_q}}
\end{align}
which satisfy $\star_L^2 = (-1)^{p(d-p)+1}$ and $\star_R^2 = (-1)^{q(d-q)+1}$. Finally, there are
adjoint derivatives $\dd_L^\dag$ and $\dd_R^\dag$ which we define by
\begin{align}
    (\dd_L^\dag A)_{\mu_2\dots\mu_p|\nu_1\dots\nu_q} &= \partial^{\mu_1} A_{\mu_1\dots\mu_p|\nu_1\dots\nu_q} \\
    (\dd_R^\dag A)_{\mu_1\dots\mu_p|\nu_2\dots\nu_q} &= \partial^{\nu_1} A_{\mu_1\dots\mu_p|\nu_1\dots\nu_q}
\end{align}
which are also nilpotent $(\dd_L^\dag)^2= (\dd_R^\dag)^2 = 0$. $N$ is $\dd_L$-coclosed if $\dd_L^\dag N = 0$ and $\dd_L$-coexact if $N=\dd_L^\dag M'$ for some \pq{p+1,q}-tensor $M'$. There are analogous definitions for $\dd_R^\dag$.

\section{The charges of the graviton}
\label{sec:Penrose_review}

We  consider the Fierz-Pauli theory of a massless spin-2 field
$h_{\mu\nu}$ propagating on a background spacetime $\M$. This is taken to be a $d$-dimensional Minkowski space, possibly with some points or regions removed so as to allow for non-trivial topology. In later sections we also consider toroidal compactifications of some spatial directions. Here we present only a summary of various conserved charges which can be associated with the graviton, following \cite{HullYetAppear,PAPER1}. The reader should refer to Appendix~\ref{app:conventions} for detailed definitions and to Ref.~\cite{PAPER1} for details.

\subsection{Penrose charges, ADM charges and magnetic charges}
\label{sec:charges_charges}

We define a linearised connection $\Gamma_{\mu\nu|\rho}$ which is a [2,1]-tensor with components
\begin{equation}
    \Gamma_{\mu\nu|\rho} = \partial_{[\mu} h_{\nu]\rho}
\end{equation}
The graviton has gauge transformations
\begin{equation}
    h_{\mu\nu} \to h_{\mu\nu} + \partial_{(\mu}\xi_{\nu)}
\end{equation}
under which the linearised Riemann tensor
\begin{equation}
    R_{\mu\nu\rho\sigma} = \partial_\rho \Gamma_{\mu\nu|\sigma} - \partial_\sigma \Gamma_{\mu\nu|\rho}
\end{equation}
is gauge-invariant. In the linearised theory, $\Gamma_{\alpha\beta|\gamma}$ is related to the linearised Christoffel connection by a gauge transformation.

The ADM charges in linearised gravity have a surface integral representation \cite{Abbott1982StabilityConstant}
\begin{equation}\label{eq:ADM_charges}
    Q[k] = \int \star J[k]
\end{equation}
where $k$ is a Minkowski space Killing vector, and $J[k]$ is the `secondary current' which satisfies
\begin{equation}\label{eq:j[k]}
    \partial^\nu J[k]_{\mu\nu} = G_{\mu\nu} k^\nu \equiv j[k]_\mu
\end{equation}
so is conserved when $R_{\mu\nu}=0$. If the field equation $G_{\mu\nu}=T_{\mu\nu}$ holds, then $J[k]$ is conserved away from sources. The explicit form of $J[k]$ is given in Appendix~\ref{app:conventions}. In regions with sources, the result is non-vanishing and gives the 1-form `primary' conserved current $j[k]$ which has been referred to as the Einstein current in Ref.~\cite{Lindstrom2022Geometrycurrents}.

As mentioned in the previous section, the ADM charges can be related to the covariant Penrose charges which are integrals of the Penrose 2-form \eqref{eq:Penrose2form}. This 2-form is constructed from a CKY tensor, which satisfies
\begin{equation}\label{eq:CKY_equation}
    \partial_\lambda K_{\mu\nu} = \tilde{K}_{\lambda\mu\nu} + 2 \eta_{\lambda[\mu}\hat{K}_{\nu]}
\end{equation}
with
\begin{equation}\label{eq:Ktilde_Khat_def_intro}
    \tilde{K}_{\lambda\mu\nu} = \partial_{[\lambda}K_{\mu\nu]} \qq{and} \hat{K}_\mu = \frac{1}{d-1} \partial^\nu K_{\nu\mu}~.
\end{equation}

For comparison with analysis which will be used later, we 
give a brief representation theory derivation of the result
 that the 2-form $K$ in the Penrose 2-form $Y[K]$ must be a CKY tensor. Demanding conservation of $Y[K]$ implies that 
\begin{equation}\label{eq:Penrose_constraint}
    0 = \partial^\mu Y[K]_{\mu\nu} = \partial^\mu (R_{\mu\nu\alpha\beta} K^{\alpha\beta} ) = R_{\mu\nu\alpha\beta} \partial^\mu K^{\alpha\beta}
\end{equation}
where we have used the contracted Bianchi identity $\partial^\mu R_{\mu\nu\alpha\beta}=0$. This follows from the Bianchi identity $\partial_{[\mu}R_{\alpha\beta]\gamma\delta}=0$ and the vacuum field equation $R_{\mu\nu}=0$. Now we must impose a condition on $\partial^\mu K^{\alpha\beta}$ such that $R_{\mu\nu\alpha\beta} \partial^\mu K^{\alpha\beta}$ vanishes. This can be done via a decomposition into irreducible representations of $SO(1,d-1)$ as follows. 

Firstly, $\partial_\mu K_{\alpha\beta}$ is a $GL(d,\mathbb{R})$ tensor in the reducible \pq{1,2} representation. This is a tensor product representation, labelled by Young tableaux $\one \otimes \two$, which can instead be written as a direct sum of irreducible $GL(d,\mathbb{R})$ representations. These can be seen as \emph{reducible} representations of $SO(1,d-1)\subset GL(d,\mathbb{R})$. Decomposing further into \emph{irreducible} representations of $SO(1,d-1)$ by separating out traces, we have
\begin{equation}
    \left( \one \otimes \two \, \right)_{GL} = \left( \,\three \oplus \twoone\, \right)_{GL} = \left( \,\three \oplus \twoone \oplus \one \right)_{SO}
\end{equation}
where the $SO(1,d-1)$ representations are traceless by definition (as well as satisfying the index symmetries in eqs.~\eqref{eq:irrep_constraint} and \eqref{eq:irrep_constraint2}). This can be written out explicitly for the tensor $\partial_\mu K_{\alpha\beta}$ as
\begin{equation}\label{eq:dK_rep_decomp}
    \partial_\mu K_{\alpha\beta} = \tilde{K}_{\mu\alpha\beta} + \left( \partial_\mu K_{\alpha\beta} - \tilde{K}_{\mu\alpha\beta} - 2\eta_{\mu[\alpha}\hat{K}_{\beta]} \right) + 2\eta_{\mu[\alpha}\hat{K}_{\beta]}
\end{equation}
Here, the first term on the right-hand side is in the $\three$ representation, the term in parentheses is in the traceless $\twoone$ representation and the final term involves a vector in the $\one$ representation. Inserting this decomposition into eq.~\eqref{eq:Penrose_constraint}, the first and last terms vanish immediately from the Bianchi identity $R_{[\mu\nu\alpha]\beta}=0$ and the field equation $R_{\mu\nu}=0$. Therefore, the result vanishes if and only if we impose that the term in parentheses in eq.~\eqref{eq:dK_rep_decomp} vanishes by itself. Indeed, this immediately gives the CKY equation~\eqref{eq:CKY_equation}. We see from this that the CKY equation can be equivalently written
\begin{equation}
    \mathcal{Y}_{\twoone}^{SO}(\partial K) = 0
\end{equation}
Arguments similar to these will be essential to our more general treatment in section~\ref{sec:general_symmetries}.
This approach to defining generalised Killing tensors by representation theory was developed and applied to a variety of tensors in Ref.~\cite{Howe2018SCKYT}.

An important result for analysing the Penrose charges is the following Minkowski space decomposition of a CKY tensor
\cite{penrose_rindler_1986,Howe2018SCKYT,Hinterbichler2023GravitySymmetries}:
\begin{equation} \label{eq:CKY_Solution}
	K_{\alpha\beta} = \A_{\alpha\beta} + \B_{[\alpha}x_{\beta]} + \C_{\alpha\beta\gamma} x^\gamma + 2x_{[\alpha}\D_{\beta]\gamma}x^\gamma + \frac{1}{2} \D_{\alpha\beta} x_\gamma x^\gamma
\end{equation}
where $\A$, $\B$, $\C$, and $\D$ are constant antisymmetric tensors. 
This is the most general form of a CKY tensor in Minkowski space.
When the Penrose charges are constructed from $K$ as in eq.~\eqref{eq:Penrose2form}, the $\B$ and $\D$ terms are related to ADM charges of the graviton, while the $\A$ and $\C$ terms are related to magnetic charges of the graviton. It follows from this decomposition that $\hat{K}$ is a Killing vector and $\tilde{K}$ is a closed CKY 3-form on Minkowski space; that is, it satisfies
\begin{equation}\label{eq:CCKY_3form}
    \partial_\mu \tilde{K}_{\alpha\beta\gamma} = \frac{3}{d-2}\eta_{\mu[\alpha} \partial^\lambda \tilde{K}_{\beta\gamma]\lambda}
\end{equation}
We therefore define a Killing vector
\begin{equation}\label{eq:k_Khat}
    k_\mu = 2(d-3) \hat{K}_\mu
\end{equation}
and a closed CKY 3-form
\begin{equation}\label{eq:lambdatilde_Ktilde}
    \tilde{\lambda}_{\mu\nu\rho} = (-1)^{d-1} (d-1) \tilde{K}_{\mu\nu\rho}
\end{equation}
with coefficients chosen for later convenience. The $\B$- and $\D$-type CKY tensors in eq.~\eqref{eq:CKY_Solution} give non-zero $k$, while the $\C$- and $\D$-type CKY tensors give non-zero $\tilde{\lambda}$. Notably, the $\A$-type CKY tensors have $k=0$ and $\tilde{\lambda}=0$.

There are also dual ADM charges which stem from certain invariances of the dual graviton field \cite{HullYetAppear}. The dual graviton, $D_{\mu_1\dots\mu_{d-3}|\nu}$, is a mixed symmetry gauge field of rank \pqi{d-3,1} which describes the same degrees of freedom as the graviton $h_{\mu\nu}$.
The Killing vectors of Minkowski space, which have a crucial role in the ADM charges \eqref{eq:ADM_charges} of the graviton, are gauge parameters which leave the graviton field invariant on all field configurations and so correspond to global symmetries. Similarly, there are certain gauge transformations which leave the dual graviton invariant on all field configurations. These are analogous to the isometries generated by Killing vectors in the graviton theory, and are similarly understood as global symmetries. 
The tensorial parameters for these can be viewed as generalised Killing tensors.

The associated charges can (sometimes) be written in terms of the graviton field and interpreted as magnetic charges for the graviton \cite{HullYetAppear}. This is most easily done with a particular choice of gauge for the graviton, so that the resulting magnetic charges found in \cite{HullYetAppear} are given in that particular gauge.

In $d=4$, the dual graviton is a symmetric tensor gauge field and its charges are somewhat distinct to the case $d>4$. Here we focus on $d>4$ and will then discuss $d=4$ in section~\ref{sec:4d_charges}.
The magnetic charges are of two types, with generalised Killing tensors denoted  $\lambda$ and $\kappa$.

In the dual graviton theory, $\lambda$ is required to be a Killing-Yano (KY) $(d-3)$-form. That is, it satisfies
\begin{equation}
    \partial_\mu \lambda_{\nu_1\dots\nu_{d-3}} = \partial_{[\mu} \lambda_{\nu_1\dots\nu_{d-3}]}
\end{equation}
It is a property of rank-$p$ KY forms that their Hodge dual is a closed CKY $(d-p)$-form \cite{Frolov2008HigherdimensionalVariables, Howe2018SCKYT}. We will denote the Hodge dual of $\lambda$ by $\tilde{\lambda}$, which is then a closed CKY 3-form (i.e. satisfies eq.~\eqref{eq:CCKY_3form}).

The other type of generalised Killing tensors are the $\kappa$ tensors. In the dual theory, these are \pqi{d-4,1}-tensors and so satisfy
\begin{equation}\label{eq:kappa_d-4,1}
    \kappa_{\mu_1\dots\mu_{d-4}|\nu} = \kappa_{[\mu_1\dots\mu_{d-4}]|\nu}
\end{equation}
and
\begin{equation}\label{eq:kappa_irred}
    \kappa_{[\mu_1\dots\mu_{d-4}|\nu]} = 0
\end{equation}
Furthermore, they satisfy a differential constraint
\begin{equation}\label{eq:dkappa=0}
    \partial_{[\sigma} \kappa_{\mu_1\dots\mu_{d-4}]|\nu}=0
\end{equation}

Charges can be constructed from these generalised Killing tensors as integrals of 2-form secondary currents $J[\lambda]$ and $J[\kappa]$, similarly to eq.~\eqref{eq:ADM_charges}. We will refer to these as $\lambda$-charges and $\kappa$-charges respectively.
The secondary current $J[\lambda]$ satisfies 
\begin{equation}\label{eq:j[l]}
    \partial^\nu J[\lambda]_{\mu\nu} = \frac{(-1)^{d-1}}{d-1} \tilde{\lambda}^{\alpha\beta\gamma} R_{\mu\alpha\beta\gamma} \equiv j[\lambda]_\mu 
\end{equation}
and so is conserved by the Bianchi identity $R_{\nu[\alpha\beta\gamma]}=0$. An explicit expression for $J[\lambda]$ is given in Appendix~\ref{app:conventions}, while $J[\kappa]$ will be discussed in section~\ref{sec:kappa_symms}. In the dual description, magnetic sources can be introduced which have the effect of sourcing the Bianchi identity, so the `primary' current $j[\lambda]$ can be non-zero \cite{HullYetAppear}. In the present work we consider only the original graviton formulation of the theory and therefore treat $R_{\nu[\alpha\beta\gamma]}=0$ as an identity.

The key result of Ref.~\cite{PAPER1} is that the Penrose 2-form is related to the secondary currents $J[k]$ and $J[\lambda]$ by 
\begin{equation}\label{eq:Y=J[k]+J[l]+dZ}
    Y[K]_{\mu\nu} = J[k]_{\mu\nu} + J[\lambda]_{\mu\nu} + \partial^\rho Z[K]_{\mu\nu\rho}
\end{equation}
where $Z[K]$ is a 3-form given by
\begin{equation}
    Z[K]^{\mu\nu\rho} = 12 K^{[\mu\nu} \Gamma\indices{^{\rho\beta]}_{|\beta}}
\end{equation}
Eq.~\eqref{eq:Y=J[k]+J[l]+dZ} has been used to give a covariant generalisation of the ADM charges, and the charges associated with the $\lambda$-invariances in Ref.~\cite{PAPER1}. 

In $d>4$, the Penrose charges associated with the $\A$- and $\C$-type CKY tensors (i.e. the KY tensors) vanish on Minkowski space \cite{PAPER1, Benedetti2023GeneralizedGravitons}, leaving only the $\B$- and $\D$-type charges which are related to the ADM charges. The $\kappa$-charges, however, are not related to the Penrose 2-form. 
One of the aims of this work is to find a manifestly gauge-invariant current which agrees with $J[\kappa]$ in the gauge used in \cite{HullYetAppear}. This would give a gauge-invariant generalisation of $J[\kappa]$ that can then be used in any gauge.

\subsection{Four dimensions}
\label{sec:4d_charges}

Four dimensions are special due to the fact that the graviton and dual graviton are both symmetric tensor gauge fields, as well as certain properties of CKY tensors. 
Since the dual graviton is also a symmetric rank-2 gauge field in $d=4$ its invariances are also given by the Killing vectors of Minkowski space. Therefore, in this case, there are no independent $\lambda$ and $\kappa$ tensors and these are only relevant for the dual graviton in $d>4$. More precisely, the $\lambda$ tensors reduce to Killing vectors when $d=4$ and the $\kappa$ tensors simply become constant 1-forms (which are also Killing vectors). So the $\lambda$- and $\kappa$-charges discussed above degenerate in four dimensions.

Some of the charges associated with invariances of the four-dimensional dual graviton can be written in terms of the graviton and expressed as a surface integral of a secondary current $\tilde{J}[\tilde{k}]$ which is given in Appendix~\ref{app:conventions}. Here $\tilde{k}$ is another Killing vector, independent of $k$, which can be written in terms of a rank-2 CKY tensor as $\tilde{k} = \star\dd{K}$ (whereas $k\propto \star\dd\star K$).
Then, in $d=4$, the Penrose 2-form is related to the secondary currents by 
\begin{equation}\label{eq:4d_Penrose}
    Y[K]_{\mu\nu} = J[k]_{\mu\nu} + \tilde{J}[\tilde{k}]_{\mu\nu} + \partial^\rho Z[K]_{\mu\nu\rho}
\end{equation}
Therefore, again in $d=4$, the Penrose charges give a gauge-invariant generalisation of the ADM and dual ADM charges. 

\subsection{Penrose 2-form as a special case of $\Omega[V]$}
\label{sec:Penrose_special_case}

We will later consider the more general family of conserved 2-forms $\Omega[V]$ introduced in eq.~\eqref{eq:Omega[V]_intro}.
At this point, we simply demonstrate that this more general set of 2-forms includes the Penrose 2-form $Y[K]$ for a specific choice of $V$. Namely, choosing $V$ to be given by\footnote{This form of $V[K]$ is closely related to the form of the Kastor-Traschen current for a KY tensor $K$ (see Appendix of Ref.~\cite{Kastor2004ConservedTensors}).}
\begin{equation} \label{eq:V_Penrose}
    V[K]\indices{_{\mu\nu\alpha\beta}^{|\gamma\delta}} = 6 \delta^{\gamma\delta}_{[\mu\nu} K_{\alpha\beta]}
\end{equation}
where $\delta^{\gamma\delta}_{\mu\nu} \equiv \delta^{[\gamma}_{[\mu} \delta^{\delta]}_{\nu]}$ and $K$ is a CKY 2-form yields 
\begin{equation}\label{eq:Omega[VPen]=Y[K]}
    \Omega[V]_{\mu\nu} = 6R\indices{^{\alpha\beta}_{\gamma\delta}} \delta^\gamma_{[\mu} \delta^\delta_\nu K_{\alpha\beta]} = 6R\indices{^{\alpha\beta}_{[\mu\nu}} K_{\alpha\beta]} =  R_{\mu\nu\alpha\beta} K^{\alpha\beta} = Y[K]_{\mu\nu}
\end{equation}
where we have used that $R_{\mu\nu}=0$ in the penultimate equality. Hence, this particular choice of $V$ reduces $\Omega[V]$ to the Penrose 2-form.

\section{$\kappa$-invariances and their associated charges}
\label{sec:kappa_symms}

As mentioned in the previous section, for $d\geq5$ the dual formulation of linearised gravity has two types of invariances corresponding to two different types of tensors, $\lambda$ and $\kappa$. We now discuss the $\kappa$ tensors in more detail. These are \pqi{d-4,1}-tensors $\kappa$ satisfying eqs.~\eqref{eq:kappa_d-4,1}, \eqref{eq:kappa_irred} and \eqref{eq:dkappa=0}.
In particular, eq.~\eqref{eq:kappa_irred} ensures the irreducibility of $\kappa$ under $GL(d,\mathbb{R})$. In the dual theory, this is essential for the dual graviton field itself to remain irreducible. The irreducibility of the dual graviton is necessary for the two descriptions of linearised gravity to describe the same number of degrees of freedom. This irreducibility implies that $\kappa$ transforms under a $GL(d,\mathbb{R})$ representation labelled by a Young tableau with two columns of lengths $d-4$ and 1.

We define the dual of a $\kappa$ tensor, which we will denote $\tilde{\kappa}$, by 
\begin{equation}
    \tilde{\kappa} = \star_L \kappa
\end{equation}
which has components
\begin{equation}\label{eq:kappa_tilde}
    \tilde{\kappa}_{\alpha\beta\gamma\delta|\nu} = \frac{1}{(d-4)!} \epsilon_{\alpha\beta\gamma\delta\mu_1\dots\mu_{d-4}} \kappa\indices{^{\mu_1\dots\mu_{d-4}}_{|\nu}}
\end{equation}
This is a \pq{4,1}-tensor,
\begin{equation}\label{eq:ktilde_4,1}
    \tilde{\kappa}_{\mu\nu\rho\sigma|\alpha} = \tilde{\kappa}_{[\mu\nu\rho\sigma]|\alpha} 
\end{equation}
which satisfies
\begin{equation}\label{eq:div_ktilde}
    \partial^\mu \tilde{\kappa}_{\mu\nu\rho\sigma|\alpha} = 0
\end{equation}
from eq.~\eqref{eq:dkappa=0}.
If $\kappa$ is irreducible, i.e.\ if it satisfies eq.~\eqref{eq:kappa_irred}, then $\tilde{\kappa}$ is traceless, i.e. $\tilde{\kappa}\indices{_{\mu\nu\rho\sigma|}^\sigma}=0$. 

With a particular gauge choice for both the graviton theory and the dual graviton theory, the connection for the dual graviton theory is dual to the connection $\Gamma$ for the graviton theory Ref.~\cite{HullYetAppear}.
For the graviton, this gauge choice is 
\begin{equation}\label{eq:gaugeg}
    \Gamma\indices{_{\mu\nu|}^\nu}=0
\end{equation}
In this gauge, the secondary current for the dual graviton theory associated with the $\kappa$-invariances can be written locally in terms of the graviton in this gauge 
for all $\kappa$ tensors as \cite{HullYetAppear}
\begin{equation}\label{eq:J[kappa]}
    J[\kappa]_{\mu\nu} = (-1)^{d} \frac{2}{d-1} \tilde{\kappa}_{\mu\nu\rho\sigma|\alpha} \Gamma^{\alpha\sigma|\rho}
\end{equation}
which satisfies
\begin{equation}\label{eq:j[kappa]}
    \partial^\nu J[\kappa]_{\mu\nu} = \frac{(-1)^{d}}{d-1} \tilde{\kappa}_{\mu\alpha\beta\gamma|\sigma} R^{\sigma\alpha\beta\gamma} \equiv j[\kappa]_\mu
\end{equation}
so is conserved by the Bianchi identity $R_{\sigma[\alpha\beta\gamma]}=0$. 
As for $j[\lambda]$ in eq.~\eqref{eq:j[l]}, the `primary' current $j[\kappa]$ can be non-zero when sources are included in the dual formulation of the theory. Here, we consider only the source-free case and so $j[\kappa]=0$ identically.
Now, integrating over a codimension-2 cycle $\Sigma_{d-2}$ gives a conserved charge~\cite{HullYetAppear}
\begin{equation}\label{eq:Q[kappa]}
    Q[\kappa] = \int_{\Sigma_{d-2}} \star J[\kappa]
\end{equation}
which we  refer to as a $\kappa$-charge.

The secondary current $J[\kappa]$ in eq.~\eqref{eq:J[kappa]} can  be written locally as a total divergence
\begin{equation}\label{eq:J[kappa]=dZ[kappa]}
    J[\kappa]_{\mu\nu} = \partial^\rho Z[\kappa]_{\mu\nu\rho}
\end{equation}
with $Z[\kappa]$  given by
\begin{equation}\label{eq:Z[kappa]}
    Z[\kappa]_{\mu\nu\rho} = (-1)^{d} \frac{1}{d-1} \, \tilde{\kappa}_{\mu\nu\rho\sigma | \alpha} h^{\sigma\alpha}
\end{equation}
This implies that the charge $Q[\kappa]$ is  topological and is non-zero only when $h_{\mu\nu}$ is a non-globally defined field configuration.
Indeed, as $\star J[\kappa]$ is closed, the charge $\int_{\Sigma_{d-2}} \star J[\kappa]$, 
for a given cycle $\Sigma_{d-2}$,
depends only on the cohomology class of $\star J[\kappa]$ and vanishes if it is an exact form. As locally
$\star J=\dd(\star Z)$, it will be exact only if $\star Z$ is a globally defined form, which will be the case if $h_{\mu\nu}$ is globally defined. See Refs.~\cite{PAPER1, HullYetAppear} for further discussion.

Note that the secondary Noether current $J[\kappa]$ derived in terms of the dual graviton in Ref.~\cite{HullYetAppear} can be written locally in terms of the graviton for any $\tilde{\kappa}$ tensors satisfying eq.~\eqref{eq:div_ktilde}, whereas $J[\lambda]$
can be written locally in terms of the graviton
if $\tilde{\lambda}$ is constant. 

We note from eq.~\eqref{eq:Z[kappa]} that $J[\kappa]$ will be conserved whether $\kappa$ is irreducible or not (i.e. whether $\tilde{\kappa}$ is traceless or not). Therefore, as far as the charges constructed with $\kappa$ are concerned, there is no need for $\kappa$ to be irreducible and charges can equally well be defined if it is reducible. 
The condition of irreducibility came from requiring the charge to result from  an invariance of the dual graviton theory.
Since we work directly in the graviton theory in this paper we will \emph{not} insist that $\kappa$ satisfy eq.~\eqref{eq:kappa_irred}. That is, here we require only that $\kappa$ is a \pq{d-4,1}-tensor (rather than an irreducible \pqi{d-4,1}-tensor). A priori, $\tilde{\kappa}$ is a \pq{4,1}-tensor. The reducible \pq{4,1} representation of $GL(d,\mathbb{R})$ is a tensor product  $\four \otimes \one = \five \oplus \fourone$. The component of $\tilde{\kappa}$ in the $\five$ representation does not contribute to the charge $Q[\kappa]$ as $Z[\kappa]$ in eq.~\eqref{eq:Z[kappa]} vanishes identically when $\tilde{\kappa}$ is fully antisymmetric. Therefore, only those $\tilde{\kappa}$-tensors in the irreducible \pqi{4,1} representation of $GL(d,\mathbb{R})$ contribute to non-zero charges, so we will restrict $\tilde{\kappa}$ to be a \pqi{4,1}-tensor.

\section{$\kappa$-charge as a special case of $\Omega[V]$}
\label{sec:kappa_subcase}
\subsection{Charges for Minkowski space}

The Penrose currents $Y[K]$, which have been shown to give a gauge-invariant construction of the ADM and $\lambda$-charges,   are not related to the $\kappa$-charges. Indeed, as we have stressed, $Y[K]$ is not the most general covariant conserved 2-form. Instead, the most general 2-form which is linear in the linearised Riemann tensor takes the form
\begin{equation}\label{eq:Omega[V]}
    \Omega[V]_{\mu\nu} = R^{\alpha\beta\gamma\delta}V_{\mu\nu|\alpha\beta|\gamma\delta}
\end{equation}
where $V$ is a \pq{2,2,2}-tensor which must satisfy  particular constraints in order for $\Omega[V]$ to be conserved. 
Each $V$ satisfying these constraints gives a co-closed 2-form which can be integrated to give a conserved charge
\begin{equation}
    \mathbb{Q}[V] = \int_{\Sigma_{d-2}} \star \Omega[V]
\end{equation}
where, $\Sigma_{d-2}$ is a codimension-2 cycle. 
If $\Omega[V]$ is co-exact, then $\mathbb{Q}[V]=0$ on any cycle $\Sigma_{d-2}$ and there is no corresponding 1-form symmetry.

We have seen in section~\ref{sec:Penrose_special_case} that $\Omega[V]$ reduces to the Penrose 2-form $Y[K]$ when $V$ is given by eq.~\eqref{eq:V_Penrose}, which then provides a covariantisation of the ADM and dual ADM charges in eq.~\eqref{eq:Y=J[k]+J[l]+dZ}. We now ask if there is a choice of $V$ for which $\Omega[V]$ agrees with the secondary current $J[\kappa]$ introduced in eq.~\eqref{eq:J[kappa]}, subject to the gauge choice~\eqref{eq:gaugeg}.
If such a construction were possible, then the charge $\mathbb{Q}[V]$ would provide a gauge-invariant formulation of the $\kappa$-charges.

Given a $\kappa$-tensor, suppose there exists a \pq{4,2}-tensor $V[\tilde{\kappa}]$ satisfying
\begin{equation}\label{eq:divV_kappa}
    \partial^\mu V[\tilde{\kappa}]_{\mu\nu\alpha\beta|\gamma\delta} = \frac{(-1)^d}{d-1} \tilde{\kappa}_{\nu\alpha\beta[\gamma|\delta]}
\end{equation}
Then, for such a  $V$, we find
\begin{align}
    \Omega[V[\tilde{\kappa}]]_{\mu\nu} &= R^{\alpha\beta\gamma\delta} V[\tilde{\kappa}]_{\mu\nu\alpha\beta|\gamma\delta} 
    = J[\kappa]_{\mu\nu} + \partial^\alpha \Phi_{\mu\nu\alpha} \label{eq:Omega_J[kappa]_link}
\end{align}
where
\begin{equation}\label{eq:Phi}
    \Phi_{\mu\nu\alpha} = 2 \Gamma^{\gamma\delta|\beta} V[\tilde{\kappa}]_{\mu\nu\alpha\beta|\gamma\delta}
\end{equation}
This implies that $\Omega[V[\tilde{\kappa}]]$ is conserved,
\begin{equation}\label{eq:conse}
    \partial ^\mu \Omega[V[\tilde{\kappa}]]_{\mu\nu} =0
\end{equation}
from eq.~\eqref{eq:J[kappa]=dZ[kappa]}, so that
\begin{equation}\label{eq:Q[kappa]_covariant}
    \mathbb{Q}[\kappa] \equiv \int_{\Sigma_{d-2}} \star \Omega[V[\tilde{\kappa}]] 
\end{equation}
is a \emph{gauge-invariant} conserved charge constructed from the $\kappa$ tensor.
Eq.~\eqref{eq:Omega_J[kappa]_link} gives an explicit relation between the secondary currents $J[\kappa]$ associated with the $\kappa$-invariances of the dual graviton and the covariant 2-forms $\Omega[V[\tilde{\kappa}]]$ so that the charges are related by
\begin{equation}\label{eq:Q[kappa]_covariantrel}
    \mathbb{Q}[\kappa] = Q[\kappa] + \int_{\Sigma_{d-2}} \dd \star \Phi
\end{equation}
Using eq.~\eqref{eq:Z[kappa]}, the covariant charge $\mathbb{Q}[\kappa]$ is a total derivative
\begin{equation}\label{eq:Q[kappa]_div}
    \mathbb{Q}[\kappa]  =  \int_{\Sigma_{d-2}} \dd \star (Z[\kappa]+\Phi)
\end{equation}
and so is a topological charge
which will only be non-zero if $h_{\mu\nu}$ is not
globally defined.

The covariant charge $\mathbb{Q}[\kappa]$ is then the sum of $Q[\kappa]=\int\dd \star Z[\kappa]$ and a further total derivative term $\int\dd \star \Phi$.
Neither $Z[\kappa]$ nor $\Phi$ is gauge-invariant, but their sum gives the gauge-invariant charge $\mathbb{Q}[\kappa]= \int\dd \star (Z[\kappa]+\Phi)$.
In section~\ref{sec:examples} we will consider examples where the charge $\mathbb{Q}[\kappa]$ is non-zero and will find a choice of gauge where the $\Phi$ term does not contribute, giving $\mathbb{Q}[\kappa] = Q[\kappa]$. In other gauges, both terms contribute to $\mathbb{Q}[\kappa]$ for these examples.
However, in general the $\Phi$ term need not vanish with the gauge condition \eqref{eq:gaugeg}, so in general the two charges $\mathbb{Q}[\kappa]$ and $Q[\kappa]$  differ, even with the gauge choice \eqref{eq:gaugeg}.
Both are conserved topological charges given by the integral of a total derivative, with $\mathbb{Q}[\kappa]$ a gauge-invariant charge and $Q[\kappa]$ a non-gauge-invariant Noether-type charge associated with a gauge invariance of the dual theory.
Recall that for branes, both covariant and non-covariant charges play a role and are related but distinct \cite{Marolf:2000cb}.

We can gain some further insight by analysing these charges in the dual formulation of the theory. There, $Q[\kappa]$ is an electric charge for the dual graviton theory, while the covariant charge $\mathbb{Q}[\kappa]$ can be understood as a covariant charge for the dual graviton. Neither $Q[\kappa]$ nor $\mathbb{Q}[\kappa]$ is a topological charge for the dual graviton theory, but their difference is the integral of a total derivative that will vanish for solutions for which the dual graviton is globally defined, but can give a non-zero topological charge more generally. Adding this topological term to $Q[\kappa]$ then covariantises it to give the gauge-invariant charge $\mathbb{Q}[\kappa]$.
We will return to this observation in an upcoming work.

This discussion hinges on the existence of a \pq{4,2}-tensor $V[\tilde{\kappa}]$ satisfying eq.~\eqref{eq:divV_kappa} for a given $\kappa$-tensor. If the background spacetime is Minkowski space, possibly with some points or regions removed but \emph{without} any compactified directions, a solution is easily found to be
\begin{equation}\label{eq:V_kappa}
    V[\tilde{\kappa}]_{\mu\nu\alpha\beta|\gamma\delta} = \frac{(-1)^d}{d-1} \tilde{\kappa}_{\mu\nu\alpha\beta|[\gamma} x_{\delta]}
\end{equation}
It is straightforward to verify that this satisfies eq.~\eqref{eq:divV_kappa} for any $\tilde{\kappa}$-tensor by using eq.~\eqref{eq:div_ktilde}. This choice of $V$ is clearly not unique, and any \pq{4,2}-tensor $V_0$ satisfying $\partial^\mu (V_0)_{\mu\nu\alpha\beta|\gamma\delta}=0$ can be added to $V[\tilde{\kappa}]$ to define another valid solution. At this point, we merely point out that a choice of $V$ is available which relates the charges $\mathbb{Q}[V]$ and $Q[\kappa]$ on Minkowski space.
In section~\ref{sec:case0} we will discuss in more generality the charges $\mathbb{Q}[V_0]$ which can arise from the inclusion of such extra terms in eq.~\eqref{eq:V_kappa}. 

In the more general discussion of section~\ref{sec:general_symmetries}, we will see that it is sufficient to require $V$ to be an irreducible \pqi{4,2}-tensor (rather than an arbitrary \pq{2,2,2}-tensor) in order to construct all the distinct charges $\mathbb{Q}[V]$. Therefore, only the irreducible \pqi{4,2} part of $V[\tilde{\kappa}]$ given in eq.~\eqref{eq:V_kappa} will contribute to the charges. Note that if $V$ is irreducible, then eq.~\eqref{eq:divV_kappa} implies that $\tilde{\kappa}$ is an irreducible \pqi{4,1}-tensor. This is not a restriction as only irreducible $\tilde{\kappa}$ tensors give non-zero charges (see section~\ref{sec:kappa_symms}).

\subsection{Choice of $V$ for solutions with compact directions}\label{sec:kappa_subcase 2}

The situation is more complicated when the background spacetime has compact directions. 
If $\M$ is the product of $T^n$ with $(d-n)$-dimensional Minkowski space (or Minkowski space with some regions removed) then $V$ must be periodic in the torus coordinates $y^i$ to be well-defined. In this work, we will achieve this by requiring that $\partial_{y^i} V_{\mu\nu\alpha\beta|\gamma\delta}=0$. We now discuss the explicit constructions which give solutions of eq.~\eqref{eq:divV_kappa} on such spaces.

Let us begin with the simplest case of $d=5$ and $n=1$, so $\M = \mathbb{R}^{1,3} \times S^1$. We denote the coordinates by $x^\mu = (x^m, y)$.
Crucially, the definition \eqref{eq:V_kappa} is not  well-defined on this space since the component
\begin{equation}\label{eq:V_non_periodic}
    V[\tilde{\kappa}]_{\mu\nu\rho\sigma|my} = -\frac{1}{4} (\tilde{\kappa}_{\mu\nu\rho\sigma|m} y - \tilde{\kappa}_{\mu\nu\rho\sigma|y} x_m)
\end{equation}
 is not periodic in $y$. 
Indeed, for a general choice of $\tilde{\kappa}$ on a background spacetime which has compact directions, eq.~\eqref{eq:divV_kappa} may not have a well-defined solution and the manipulations leading to eq.~\eqref{eq:Omega_J[kappa]_link} may break down. 

However, for the five-dimensional case, there are two possible solutions to this apparent problem which can arise for specific choices of $\tilde{\kappa}$. The first possibility is the simple observation that $V[\tilde{\kappa}]_{\mu\nu\rho\sigma|my}$ in eq.~\eqref{eq:V_non_periodic} above \emph{is} well-defined if the components $\tilde{\kappa}_{\mu\nu\rho\sigma|m}=0$. That is, for particular choices of $\tilde{\kappa}$, constructing $V[\tilde{\kappa}]$ as in eq.~\eqref{eq:V_kappa} remains valid. 
The second possibility is that, while $V[\tilde{\kappa}]$ in eq.~\eqref{eq:V_kappa} may be ill-defined for a given $\tilde{\kappa}$, there may exist another \pqi{4,2}-tensor $V$ which solves eq.~\eqref{eq:divV_kappa}. Then the manipulations leading to the crucial result in eq.~\eqref{eq:Omega_J[kappa]_link} go through as before. The expression given in eq.~\eqref{eq:V_kappa} is merely one possible solution on Minkowski space. 
To summarise, for a given solution there may not be a $V$ satisfying~\eqref{eq:divV_kappa} for general $\kappa$, so we seek those $\kappa$ for which there is such a $V$ and then use that $V[\kappa]$ in our construction.
 
Let us study a pair of cases that are relevant to the examples we consider later. In $d=5$, the irreducible $\kappa$ tensors are in the $\oneone$ representation.
For the example studied in section~\ref{sec:KKmono}, the $\kappa$-tensor that gives a non-zero charge has non-vanishing components 
\begin{equation}\label{eq:first_example_kappaa}
    \kappa_{y|y} = \kappa_0
\end{equation}
where $\kappa_0$ is a constant. 
The dual \eqref{eq:kappa_tilde} of this tensor has non-vanishing components
\begin{equation}\label{eq:first_example_kappa}
    \tilde{\kappa}_{tabc|y} = \kappa_0 \epsilon_{abc}
\end{equation}
Note that, in particular, the $\tilde{\kappa}_{\mu\nu\rho\sigma|m}$ components all vanish and so $V[\tilde{\kappa}]$ given  by eq.~\eqref{eq:V_kappa} is well-defined. This is a realisation of the first solution outlined above.

For the charge studied in section~\ref{sec:TaubNUT}, the relevant  $\kappa$-tensor has non-zero components 
\begin{equation}\label{eq:second_example_kappaa}
    \kappa_{t|y} = \kappa_{y|t} = \kappa_0
\end{equation}
The dual of this tensor has
\begin{equation}\label{eq:second_example_kappa}
    \tilde{\kappa}_{abct|t} = \tilde{\kappa}_{abcy|y} = -\epsilon_{abc} \kappa_0
\end{equation}
and all other components of $\tilde{\kappa}$ which are not related to these by anti-symmetry vanish.
In this case, $V[\tilde{\kappa}]$ defined in eq.~\eqref{eq:V_kappa} is not well-defined on the compact direction and we must seek a different, well-defined $V$ which satisfies eq.~\eqref{eq:divV_kappa} for this $\tilde{\kappa}$. 
We consider instead the \pqi{4,2}-tensor $V$ with components 
\begin{equation}\label{eq:second_example_V}
    V_{abct|dt} = V_{abcy|dy} = -\frac{1}{8} \kappa_0 \epsilon_{abc} x_d
\end{equation}
and all other components not related to these by anti-symmetry vanish. It can be straightforwardly checked that this solves eq.~\eqref{eq:divV_kappa} for $\tilde{\kappa}$ given in eq.~\eqref{eq:second_example_kappa}. Furthermore, it has no dependence on the $S^1$ coordinate $y$, so is well-defined on $\M$.
It is a \pqi{4,2}-tensor as $\epsilon_{[abc} x_{d]}=0$ with $a,b=1,2,3$. This is a realisation of the second solution described above.

We now consider solutions in $d>5$ dimensions. In particular, we fix $\M =\mathbb{R}^{1,3} \times T^n$ with $n>1$. Cases with Minkowski space of greater than four dimensions can be analysed in an analogous manner. With $n>1$, $V[\tilde{\kappa}]$ in eq.~\eqref{eq:V_kappa} has components
\begin{align}
    V[\tilde{\kappa}]_{\mu\nu\rho\sigma|mn} & = \frac{(-1)^d}{2(d-1)} \left( \tilde{\kappa}_{\mu\nu\rho\sigma|m} x_n - \tilde{\kappa}_{\mu\nu\rho\sigma|n} x_m \right) \label{eq:V_kappa_mn} \\
    V[\tilde{\kappa}]_{\mu\nu\rho\sigma|mi} &= \frac{(-1)^d}{2(d-1)} \left( \tilde{\kappa}_{\mu\nu\rho\sigma|m} y_i - \tilde{\kappa}_{\mu\nu\rho\sigma|i} x_m \right) \label{eq:V_kappa_mi} \\
    V[\tilde{\kappa}]_{\mu\nu\rho\sigma|ij} &= \frac{(-1)^d}{2(d-1)} \left( \tilde{\kappa}_{\mu\nu\rho\sigma|i} y_j - \tilde{\kappa}_{\mu\nu\rho\sigma|j} y_i \right) \label{eq:V_kappa_ij}
\end{align}
In order for the $V[\tilde{\kappa}]_{\mu\nu\rho\sigma|mi}$ components to be well-defined on the toroidal directions, we require $\tilde{\kappa}_{\mu\nu\rho\sigma|m}=0$ for all $m$. For the $V[\tilde{\kappa}]_{\mu\nu\rho\sigma|ij}$ components to be well-defined, we require $\tilde{\kappa}_{\mu\nu\rho\sigma|i}=0$. Therefore, unless $\tilde{\kappa}=0$ (and so $V[\tilde{\kappa}]=0$), this choice of \pq{4,2}-tensor $V[\tilde{\kappa}]$ is not well-defined. Therefore, we are forced to find an alternative solution to eq.~\eqref{eq:divV_kappa}. 

We now  describe such solutions for the  $\tilde{\kappa}$-tensors given above in eqs.~\eqref{eq:first_example_kappa} and \eqref{eq:second_example_kappa}. 
In $d>5$ dimensions, irreducible $\kappa$-tensors are in the \pqi{d-4,1} representation. The $\kappa$-tensor whose dual is the $\tilde{\kappa}$-tensor in eq.~\eqref{eq:first_example_kappa} has components
\begin{equation}\label{eq:first_kappa_higher_dim}
    \kappa_{y_1 y_2 \dots y_{d-4}|y_1} = \kappa_0 
\end{equation}
in $d$-dimensions. Let us define a \pq{4,2}-tensor $V$ with components 
\begin{equation}\label{eq:V_first_case}
    V_{\mu\nu\rho\sigma|mi} = V[\tilde{\kappa}]_{\mu\nu\rho\sigma|mi}
\end{equation}
with the right-hand side given in eq.~\eqref{eq:V_kappa_mi} and all other components vanishing. Since $\tilde{\kappa}_{\mu\nu\rho\sigma|m}=0$ for this example, this $V$ is well-defined. It is straightforward to verify that it satisfies eq.~\eqref{eq:divV_kappa} for $\tilde{\kappa}$ in eq.~\eqref{eq:first_example_kappa}. 

Finally, let us the consider the $d>5$ dimensional case with $\tilde{\kappa}$ given by eq.~\eqref{eq:second_example_kappa}. The $\kappa$-tensor dual to this is
\begin{equation}\label{eq:second_kappa_higher_dim}
    \kappa_{ty_2\dots y_{d-4}|y_1} = \kappa_{y_1 y_2\dots y_{d-4}|t} = \kappa_0
\end{equation}
The tensor $V$ constructed in eq.~\eqref{eq:second_example_V} remains well-defined and satisfies eq.~\eqref{eq:divV_kappa} for this $\tilde{\kappa}$.

\subsection{Condition for non-trivial charge}
\label{sec:kappa_needs_cohomology}

While a conserved current $\Omega[V]$ can be built for each \pq{4,1}-tensor $\tilde{\kappa}$ by using $V[\kappa]$ in eq.~\eqref{eq:V_kappa}, we note that if $\tilde{\kappa}$ is of the form
\begin{equation}
\label{eq:kappa=drho}
    \tilde{\kappa}_{\mu\nu\alpha\beta|\gamma} = \partial^\sigma \rho_{\sigma\mu\nu\alpha\beta|\gamma}
\end{equation}
for some \pq{5,1}-tensor $\rho$, then the charge $\mathbb{Q}[\kappa]$ vanishes. This follows as 
\begin{equation}\label{eq:surfa}
    \Omega[V[\tilde{\kappa}]]_{\mu\nu} = \frac{(-1)^d}{d-1} R^{\alpha\beta\gamma\delta} \partial^\sigma \rho_{\sigma\mu\nu\alpha\beta|\gamma} x_\delta = \frac{(-1)^d}{d-1} \partial^\sigma (R^{\alpha\beta\gamma\delta} \rho_{\sigma\mu\nu\alpha\beta|\gamma} x_\delta) \equiv \partial^\sigma \Xi_{\mu\nu\sigma}
\end{equation}
where we have used $\partial_{[\sigma}R_{\alpha\beta]\gamma\delta}=0$ in the final equality. $\Xi$ is a globally defined 3-form since it depends only on the curvature, so $\mathbb{Q}[\kappa] = \int \star\Omega[V[\tilde{\kappa}]] = \int \dd\star \Xi = 0$ by Stokes' theorem as the integral is assumed to be taken over a cycle. Therefore, non-trivial $\mathbb{Q}[\kappa]$ charges only occur when the spacetime admits $\tilde{\kappa}$-tensors satisfying eq.~\eqref{eq:div_ktilde} which are not of the form \eqref{eq:kappa=drho}. This implies a notion of cohomology which we will develop in the following sections.

\section{Charges and cohomology}
\label{sec:kappa_cohomology}

We have seen in section~\ref{sec:kappa_needs_cohomology} that if $\kappa$ is $\dd_L$-exact, then the charge $\mathbb{Q}[\kappa]$ vanishes. Indeed, the same is true of the non-covariant charge $Q[\kappa]$ \cite{HullYetAppear}. Hence, in order to construct non-zero charges we require a $\kappa$-tensor in a non-trivial $\dd_L$-cohomology class. 

Furthermore, in order for the charge $\mathbb{Q}[\kappa] = \int\star\Omega[V[\tilde{\kappa}]]$ to be non-zero, the $(d-2)$-form $\star\Omega[V]$ must itself 
be closed and not exact, so that it
represents a non-trivial de Rham cohomology class. There are, therefore, two different forms which must be in non-trivial cohomology classes for there to be a non-zero charge. We consider them in turn below.

\subsection{$\dd_L$-cohomology for $\kappa$-tensors}

We now consider the general form for the $\kappa$-tensors, following \cite{HullYetAppear}.
The $\kappa$-tensors satisfy
eq.~\eqref{eq:dkappa=0}, which we repeat here:
\begin{equation}\label{eq:kappa_constraint}
    \partial_{[\mu}\kappa_{\alpha_1\dots\alpha_{d-4}]|\beta} = 0
\end{equation}
i.e.\ they are $\dd_L$-closed.
As pointed out previously, $\dd_L$-exact $\kappa$-tensors will satisfy this automatically. That is, if $\kappa_{\alpha_1\dots\alpha_{d-4}|\beta} = \partial_{[\alpha_1}\rho_{\alpha_2\dots\alpha_{d-4}]|\beta}$ for some \pq{d-5,1}-tensor $\rho$, then eq.~\eqref{eq:kappa_constraint} is trivially satisfied.\footnote{In $d=5$ this remains true when $\kappa$ is reducible, but not when it is irreducible. In that case, the relevant $\dd_L$-exact form for a $\kappa$-tensor is $\kappa_{\mu|\nu} = \partial_\mu \partial_\nu \psi$ for a scalar $\psi$.} We also know from section~\ref{sec:kappa_needs_cohomology} that the charges $Q[\kappa]$ and $\mathbb{Q}[\kappa]$ vanish in this case, so these charges depend only on the 
$\dd_L$-cohomology class.

We begin with the case of reducible $\kappa$-tensors, which are in the reducible \pq{d-4,1} representation. This is a tensor product representation of the fully antisymmetric $(d-4)$-form representation and the 1-form representation.
As such, reducible $\kappa$-tensors can be written as a linear combination of tensors of the basic type
\begin{equation}\label{eq:kappa_red}
    \kappa^{\text{red.}}_{\alpha_1\dots\alpha_{d-4}|\beta} = \tau_{\alpha_1\dots\alpha_{d-4}} \sigma_\beta
\end{equation}
where $\tau$ is a $(d-4)$-form and $\sigma$ is a 1-form. 
For such a tensor, eq.~\eqref{eq:kappa_constraint} is satisfied if $\tau$ is closed and $\sigma$ is constant, and will not be $\dd_L$-exact only if $\tau$ is not exact.
Then the non-trivial $\kappa$-tensors are constructed from $(d-4)$-forms
$\tau$ that represent a non-trivial cohomology class, together with constant 1-forms.  Requiring non-trivial $\dd_L$-cohomology for $\kappa$ then relies on having a non-trivial de Rham cohomology group $H^{d-4}(\M)$.
General solutions then arise from taking linear combinations of terms of the form \eqref{eq:kappa_red}.
An alternative way of seeing this is to think of $\kappa_{\alpha_1\dots\alpha_{d-4}|\beta}$ as being a vector-valued $(d-4)$-form, which is closed and non-exact for each value of $\beta$.

If we instead consider irreducible $\kappa$-tensors, for which
\begin{equation}
    \kappa^{\text{irred.}}_{[\alpha_1\dots\alpha_{d-4}|\beta]}=0
\end{equation}
these can be written as a linear combination of tensors of the form
\begin{equation}\label{eq:irreducible_form}
    \kappa^{\text{irred.}}_{\alpha_1\dots\alpha_{d-4}|\beta} = \tau_{\alpha_1\dots\alpha_{d-4}}\sigma_\beta - \tau_{[\alpha_1\dots\alpha_{d-4}}\sigma_{\beta]} 
\end{equation}
This follows from the fact that all irreducible \pqi{p,q}-tensors can be written as the Young symmetrised version of a reducible \pq{p,q}-tensor, and because the Young symmetriser is a linear operator. Indeed, $\kappa^{\text{irred.}}$ is simply the Young symmetrised version of $\kappa^{\text{red.}}$ onto the irreducible \pqi{d-4,1} $GL(d,\mathbb{R})$ representation.

For closed $\tau$ and constant $\sigma$, we have
\begin{equation}
    \partial_{[\mu} \kappa^{\text{irred.}}_{\alpha_1\dots\alpha_{d-4}]|\beta} =  (-1)^{d-3} \frac{d-4}{d-3} \partial_{[\mu|} \tau_{\beta|\alpha_1\dots\alpha_{d-5}}\sigma_{\alpha_{d-4}]}
\end{equation}
Demanding that this vanishes implies
\begin{equation}
    \partial_\beta (\tau_{[\alpha_1\dots\alpha_{d-4}}\sigma_{\mu]}) = 0
\end{equation}
so $\tau\wedge\sigma = \text{constant}$. This is solved by
\begin{equation}
    \tau = \tau_0 + \beta \wedge \sigma
\end{equation}
where $\tau_0$ is a constant $(d-4)$-form and $\beta$ is a closed $(d-5)$-form (such that $\tau$ is closed). Substituting back into eq.~\eqref{eq:irreducible_form}, we see that
\begin{equation}\label{eq:irreducible_form_final}
    \kappa^{\text{irred.}}_{\alpha_1\dots\alpha_{d-4}|\beta} = (\kappa_0)_{\alpha_1\dots\alpha_{d-4}|\beta} + \beta_{[\alpha_1\dots\alpha_{d-5}}\sigma_{\alpha_{d-4}]}\sigma_\beta
\end{equation}
where $(\kappa_0)_{\alpha_1\dots\alpha_{d-4}|\beta} = (\tau_0)_{\alpha_1\dots\alpha_{d-4}} \sigma_\beta - (\tau_0)_{[\alpha_1\dots\alpha_{d-4}} \sigma_{\beta]}$ is a constant irreducible \pqi{d-4,1} tensor. 
A general irreducible $\kappa$ tensor can be written as a linear combination of tensors of the form of eq.~\eqref{eq:irreducible_form_final}. Since the constant term $\kappa_0$ can be written as a linear combination of tensors of the form $\beta_{[\alpha_1\dots\alpha_{d-5}}\sigma_{\alpha_{d-4}]}\sigma_\beta$ where $\beta$ is taken to be a constant $(d-5)$-form, we can equivalently write the general solution as a linear combination of tensors of the form
\begin{equation}
    \kappa^{\text{irred.}}_{\alpha_1\dots\alpha_{d-4}|\beta} = \beta_{[\alpha_1\dots\alpha_{d-5}}S_{\alpha_{d-4}]|\beta}
\end{equation}
where $S_{\alpha|\beta}$ is a symmetric constant \pqi{1,1}-tensor. Then $\kappa$ will be $\dd_L$-closed and not $\dd_L$-exact if $\beta$ is closed and not exact.

Given this knowledge of the $\kappa$ tensors, we return to the question of their $\dd_L$-cohomology. In order to construct non-trivial charges $\mathbb{Q}[\kappa]$ there must exist a $\dd_L$-closed $\kappa$ tensor which is not $\dd_L$-exact. From the considerations above, it is clear that this is only possible if the de Rham cohomology group $H^{d-4}(\M)$ is non-trivial in the case of reducible $\kappa$ tensors, and $H^{d-5}(\M)$ is non-trivial for irreducible ones. 

Let us consider the $\kappa$-tensors given in section~\ref{sec:kappa_subcase 2}, which will be used in the examples studied in section~\ref{sec:examples}. The background spacetime is $\mathbb{R}^{1,3} \times S^1$ with $y$ the coordinate on the $S^1$. Firstly, consider the five-dimensional $\kappa$-tensor in eq.~\eqref{eq:first_example_kappaa} with $\kappa_{y|y} = \kappa_0$. Let us first treat this as an irreducible tensor. If the $y$ coordinate were non-compact, then this would be trivial since $\kappa_{\mu|\nu}=\partial_\mu \partial_\nu \psi$ for $\psi = \frac{1}{2} \kappa_0 y^2$. However, for $y$ periodic, $\psi = \frac{1}{2} \kappa_0 y^2$ is not a well-defined 0-form and so in this case $\kappa$ is $d_L$-closed but not $d_L$-exact. If we treat this $\kappa$ as a reducible tensor instead, we could write $\kappa_{\mu|\nu} = \partial_\mu \xi_\nu$ with $\xi_y = \kappa_0 y$ if $y$ were a non-compact coordinate. Again, when $y$ is periodic this is not a well-defined 1-form and the $\kappa$-tensor is non-trivial.

The other example of interest uses the $\kappa$-tensor \eqref{eq:second_example_kappaa}, with components $\kappa_{y|t} = \kappa_{t|y} = \kappa_0$. Again, if $y$ were non-compact we could write $\kappa_{\mu|\nu} = \partial_\mu \partial_\nu \psi$ with $\psi = \kappa_0 ty$. However, with $y$ periodic this is not well-defined. Therefore, this $\kappa$-tensor is $\dd_L$-closed and not $\dd_L$-exact. Treating $\kappa$ as a reducible \pq{1,1}-tensor, we could write $\kappa_{\mu|\nu} = \partial_\mu \xi_\nu$ where $\xi$ has components $\xi_t = \kappa_0 y$ and $\xi_y = \kappa_0 t$. Again, this is not well-defined when $y$ is periodic, so the $\kappa$-tensor is not $\dd_L$-exact. Similarly, the higher-dimensional $\kappa$-tensors in eqs.~\eqref{eq:first_kappa_higher_dim} and \eqref{eq:second_kappa_higher_dim} are $\dd_L$-closed and not $\dd_L$-exact when there are compact directions.

In this work we study the graviton theory on Minkowski space with points or regions removed, and possibly with directions toroidally compactified. Let us first consider the case without toroidal directions. 
Minkowski space has trivial cohomology, so non-trivial charges can only arise if some region is excised from Minkowski space.
A non-trivial $H^{d-4}(\M)$ can be generated in $\mathbb{R}^{1,d-1}$ by excising a 2-brane (with 3-dimensional worldvolume) and so considering $\M = \mathbb{R} \times (\mathbb{R}^{d-1} \setminus \mathbb{R}^2)$ then $H^{d-4}(\M) = H^{d-4}(\mathbb{R}^{d-3}\setminus \{\text{point}\}) \cong \mathbb{R}$. A closed but non-exact $(d-4)$-form in $\M$ is given, for example, by the volume form of a $(d-4)$-sphere wrapping the 2-brane.
Similarly, excising a 3-brane (with 4-dimensional worldvolume) gives non-trivial $H^{d-5}(\M)$.

Consider now the case of a toroidal compactification.
Then $\M=\mathbb{R}^{1,n}\times T^m$ has non-trivial $H^{d-4}(\M)$ provided $n\leq3$ and has non-trivial $H^{d-5}(\M)$ provided $n\leq4$. Obviously, points and/or regions could also be removed from these compactified spaces to generate further non-trivial cohomology groups.

\subsection{Cohomology for $\Omega[V]$}

To obtain a non-zero charge, the $(d-2)$-form $\star\Omega[V]$ should be in non-trivial cohomology class, so $H^{d-2}(\M)$
should be non-trivial.
In the examples we consider in the next section $\M=(\mathbb{R}^{3}\setminus \{\text{point}\})
\times \mathbb{R} \times T^{d-4}$, i.e.\ the product of
a spatial $\mathbb{R}^{3}$ with a point removed
with a $(d-4)$-torus and a time-like line.
The $\mathbb{R}^{3}$ with a point removed has a non-trivial 2-cycle given by a 2-sphere surrounding the point and taking the product of this with $T^{d-4}$ gives a non-trivial cycle $\Sigma_{d-2}$ of codimension 2 that can be integrated over.
This space also has non-trivial $H^{d-5}(\M)$ and $H^{d-4}(\M)$ represented by forms on the torus and these can be used in constructing $\kappa$.

\section{Some solutions and their charges}
\label{sec:examples}

Ref.~\cite{HullYetAppear} discussed various solutions of the linearised theory which carry the charges $Q[k]$, $Q[\kappa]$ and $Q[\lambda]$ and the covariant Penrose charges of these were calculated in~\cite{PAPER1}. Here, we focus on the charges $\mathbb{Q}[\kappa]$, related to the non-covariant charges $Q[\kappa]$ by eq.~\eqref{eq:Q[kappa]_covariant}, and study solutions which carry these. We first analyse a solution of the linearised theory derived from the Kaluza-Klein monopole solution and then one obtained from Lorentzian Taub-NUT space.
 
\subsection{Gravitational monopoles in $d\geq5$}
\label{sec:KKmono}

The linearisation of the Kaluza-Klein monopole solution considered in~\cite{HullYetAppear} is a linear superposition of a solution with mass and a solution which carries a magnetic charge. Here we focus on the latter part of the solution, which is in itself a valid solution of the linearised theory. (The other part has  
$Q[\kappa]=0$ and $\mathbb{Q}[\kappa]=0$.)

We first consider the linearised Kaluza-Klein Ansatz on the space $\mathbb{R}^{1,3}\times S^1$ with the cylinder $\mathbb{R} \times S^1$ at $\vec{0}\in\mathbb{R}^3$ removed; we denote this space $(\mathbb{R}^{1,3}\times S^1)^{\bullet}$.
The coordinates will be taken to be $(t,x^a,y)$ where $y\sim y+2\pi R_y$. The metric on the space is $\eta = (-1,1,1,1,1)$. We take the graviton Ansatz
\begin{equation}\label{eq:KK_ansatz}
    h_{ay} = 2A_a(x^b)
\end{equation}
where the field strength $F = \dd{A}$ is given by
\begin{equation}
    F_{ab} = \epsilon_{abc}\partial_c \widetilde{W}
\end{equation}
and $\widetilde{W}$ is given by
\begin{equation}
    \widetilde{W}(x^a) = - \frac{p}{\abs{x}}
\end{equation}
where $p$ is a constant parameterising the monopole charge. The non-vanishing connection coefficients are
\begin{equation}
    \Gamma_{ab|y} = F_{ab} \qc \Gamma_{ay|b} = \partial_a A_b
\end{equation}
and the non-vanishing components of the linearised Riemann tensor are
\begin{equation}
    R_{abcy} = \partial_c F_{ab}
\end{equation}
It has been shown in Ref.~\cite{HullYetAppear} that this solution has a non-vanishing charge $Q[\kappa]$ corresponding to a $\kappa$-tensor with non-zero components 
\begin{equation}\label{eq:KK_example_kappa}
    \kappa_{y|y}=\kappa_0
\end{equation}
in five dimensions, with $\kappa_0$ a constant. As discussed in section~\ref{sec:kappa_cohomology}, this is a $\dd_L$-closed but not $\dd_L$-exact \pqi{1,1}-tensor on $(\mathbb{R}^{1,3}\times S^1)^\bullet$. The charge is defined over a codimension-2 surface $\Sigma_3 = S^2 \times S^1$ where $S^2$ is a 2-sphere containing the origin within a constant time hypersurface of $\mathbb{R}^{1,3}$, and $S^1$ is the circle parameterised by $y$, for which eq.~\eqref{eq:Q[kappa]} yields
\begin{equation}\label{eq:KK_Q[kappa]}
    Q[\kappa] = 8\pi^2 R_y \kappa_0 p
\end{equation}

We now evaluate the covariant charge $\mathbb{Q}[\kappa]$ given in eq.~\eqref{eq:Q[kappa]_covariant} for this solution. First note that the Hodge dual $\tilde{\kappa} = \star_L \kappa$ of the $\kappa$-tensor in eq.~\eqref{eq:KK_example_kappa}, given by eq.~\eqref{eq:kappa_tilde}, has non-vanishing components
\begin{equation}\label{eq:KK_kappatilde}
    \tilde{\kappa}_{tabc|y} = \kappa_0 \epsilon_{abc}
\end{equation}
This  $\kappa$-tensor was discussed in section~\ref{sec:kappa_subcase 2}, where we showed that the \pq{4,2}-tensor $V[\tilde{\kappa}]$ in eq.~\eqref{eq:V_kappa} is well-defined in this case.
For the  $\tilde{\kappa}$-tensor in eq.~\eqref{eq:KK_kappatilde}, $V$ has non-vanishing components
\begin{equation}
    V_{tabc|y\alpha} = -\frac{1}{8} \kappa_0 \epsilon_{abc} x_\alpha
\end{equation}
where $x^\alpha = (t,x^a)$, which is well-defined on $(\mathbb{R}^{1,3}\times S^1)^\bullet$.
Explicit calculation then yields
\begin{equation}
    \Omega[V]_{at} = p \kappa_0 \frac{x_a}{\abs{x}^3}
\end{equation}
The charge $\mathbb{Q}[\kappa]$ is defined over a codimension-2 surface which we again take as $\Sigma_3 = S^2 \times S^1$. The integration yields
\begin{equation}\label{eq:KK_cov_charge}
    \mathbb{Q}[\kappa] = \int_{\Sigma_3} \star \Omega[V] = 16 \pi^2 R_y p \kappa_0
\end{equation}

Note that this is a factor of 2 larger than $Q[\kappa]$ in eq.~\eqref{eq:KK_Q[kappa]}. The difference comes from the final term in eq.~\eqref{eq:Q[kappa]_covariant}. This solution has a non-globally defined $h$ and so $\int \dd\star\Phi$, i.e. the final term in eq.~\eqref{eq:Q[kappa]_covariant}, can be non-zero. Indeed, explicit evaluation gives
\begin{equation}\label{eq:KK_excess}
    \int_{\Sigma_3} \dd\star\Phi = 8\pi^2 R_y \kappa_0 p
\end{equation}
so that eqs.~\eqref{eq:KK_Q[kappa]} and \eqref{eq:KK_excess} contribute equally and sum to give $\mathbb{Q}[\kappa]$ in eq.~\eqref{eq:KK_cov_charge}, as expected from the general result in eq.~\eqref{eq:Q[kappa]_covariant}.

As described in section~\ref{sec:kappa_subcase}, while $\mathbb{Q}[\kappa]$ is gauge-invariant, neither of the two terms on the right-hand side of eq.~\eqref{eq:Q[kappa]_covariant} (that is, $Q[\kappa]$ and $\int\dd\star\Phi$) are gauge-invariant. In the gauge presented above (i.e. using the graviton components \eqref{eq:KK_ansatz}), the two contributions to $\mathbb{Q}[\kappa]$ given in eqs.~\eqref{eq:KK_Q[kappa]} and \eqref{eq:KK_excess} are equal. In another gauge the two individual contributions will change, with only their sum remaining invariant. 

We now ask if it is possible to find a gauge choice where $\mathbb{Q}[\kappa] = Q[\kappa]$ and $\int\dd\star\Phi=0$ for this solution, such that $\mathbb{Q}[\kappa]$ gives a gauge-invariant quantity which matches $Q[\kappa]$ in this gauge. As indicated in section~\ref{sec:kappa_subcase}, we will see that such a gauge choice does exist.
Consider a gauge transformation of the graviton field $h_{\mu\nu}$ in eq.~\eqref{eq:KK_ansatz} to
\begin{equation}
    h'_{\mu\nu} = h_{\mu\nu} + \partial_{(\mu}\xi_{\nu)}
\end{equation}
where the non-zero components of $\xi$ are taken to be
\begin{equation}\label{eq:gauge_transf_KK}
    \xi_a = 4yA_a
\end{equation}
We note that the gauge parameter $\xi$ is not globally-defined. This is, of course, not an issue. The gauge transformations need only be well-defined within each chart. This is akin to a large gauge transformation winding around the fibre direction in Maxwell theory.

The components of the gauge transformed graviton $h'_{\mu\nu}$ are
\begin{equation}
    h'_{ay} = 4A_a\qc h'_{ab} = 4y \partial_{(a} A_{b)}
\end{equation}
One can then explicitly check that the linearised Riemann tensors of the two gravitons are equal, $R_{\mu\nu\rho\sigma} = R'_{\mu\nu\rho\sigma}$, as was guaranteed. Therefore the charge $\mathbb{Q}[\kappa]$ is invariant under the gauge transformation. However, calculating the contribution from $\int\dd\star\Phi$ in this gauge gives a vanishing result. Indeed, an explicit calculation of $Q[\kappa]'$ (that is, the charge $Q[\kappa]$ evaluated for the gauge transformed graviton $h'_{\mu\nu}$) verifies that, in this gauge,
\begin{equation}\label{eq:Q=Q_gauge}
    \mathbb{Q}[\kappa] = Q[\kappa]'
\end{equation}

As discussed in section~\ref{sec:kappa_subcase}, the gravitational duality can be imposed at the level of the connection $\Gamma$ in the gauge in which $\Gamma\indices{_{\mu\nu|}^\mu}=0$. The transformation \eqref{eq:gauge_transf_KK} ensures that $h'_{\mu\nu}$ satisfies this gauge condition. We note that eq.~\eqref{eq:Q=Q_gauge} does not hold in this gauge when evaluated on arbitrary solutions, however it does hold for this particular solution.

While this solution is a magnetically charged configuration from the perspective of the graviton field, for the dual graviton description it is an electric charge. Therefore, in the dual theory, the contribution $\int\dd\star\Phi$ will vanish and eq.~\eqref{eq:Q=Q_gauge} follows. 
This would then imply that $\int\dd\star\Phi$ should also vanish for the graviton theory in this gauge, as we have seen directly.

\paragraph{Higher-dimensional solutions.} 

One can similarly construct the $\kappa$-charges associated with higher-dimensional solutions with a background spacetime $(\mathbb{R}^{1,3} \times T^{d-4})^\bullet$ for $d>5$. Let the coordinates on $T^{d-4}$ be $y^i \sim y^i + 2\pi R_i$, with $i=1,\dots,d-4$. We  take the Ansatz \eqref{eq:KK_ansatz} but with $y$ taken to be $y^1$.
In $d$ dimensions, $\kappa$ is a \pq{d-4,1}-tensor. We take its non-zero components to be
\begin{equation}\label{eq:KK_kappa_higher_dim}
    \kappa_{y_1 y_2\dots y_{d-4}|y_1} = \kappa_0 
\end{equation}
 This is proportional to the volume form of $T^{d-4}$.
This is of the form \eqref{eq:irreducible_form_final} where $\beta$ is the volume form on the $T^{d-5} \subset T^{d-4}$ spanned by $y_2,\dots,y_{d-4}$, and $\sigma=dy^1$.
This $\kappa$ is $\dd_L$-closed and not $\dd_L$-exact. Taking its $\star_L$ Hodge dual leads to the same components as in eq.~\eqref{eq:KK_kappatilde}.
As discussed in section~\ref{sec:kappa_subcase 2}, in $d>5$, $V[\tilde{\kappa}]$ in eq.~\eqref{eq:V_kappa} is not well-defined for this solution. Instead, we use the \pq{4,2}-tensor $V$ in eq.~\eqref{eq:V_first_case}. This is well-defined on $\M$ and solves eq.~\eqref{eq:divV_kappa} for the $\kappa$ in eq.~\eqref{eq:KK_kappa_higher_dim}.
The codimension-2 integration surface is now taken to be $\Sigma_{d-2} = S^2 \times T^{d-4}$ with the $S^2$ surrounding the origin of a constant time hypersurface of $\mathbb{R}^{1,3}$. The resulting charge is
\begin{equation}
    \mathbb{Q}[\kappa] = 8\pi \kappa_0 p \, \text{Vol}_{T^{d-4}}
\end{equation}
where $\text{Vol}_{T^{d-4}} = \prod_{i=1}^{d-4} 2\pi R_i$ is the volume of the $(d-4)$-torus. Indeed, this reduces to eq.~\eqref{eq:KK_cov_charge} for $d=5$.

\subsection{Lorentzian Taub-NUT in $d\geq5$}
\label{sec:TaubNUT}

Another example which carries the charge $Q[\kappa]$ is    the product of linearised Lorentzian Taub-NUT with a torus. The full Taub-NUT metric is dyonic, but here we consider a linearisation only of the magnetically charged part of the solution. In the linearised theory, this is a valid solution in its own right.
We begin with the case of $d=5$ on the background $(\mathbb{R}^{1,3}\times S^1)^\bullet$ as in the previous section. We take a graviton field with components
\begin{equation}\label{eq:TaubNUT_metric}
    h_{at} = 2A_a(x^b)
\end{equation}
where $A_a(x^b)$ satisfies
\begin{equation}\label{eq:F=*dX}
    F_{ab} = \epsilon_{abc} \partial_c X
\end{equation}
with
\begin{equation}
    X(x^a) = -\frac{N}{\abs{x}}
\end{equation}
where $N$ is a constant that quantifies the NUT charge of the field configuration.
The non-zero components of the linearised Riemann tensor are
\begin{equation}
    R_{abct} = \partial_c F_{ab}
\end{equation}
This example has been studied in Ref.~\cite{HullYetAppear}, where it is shown that it carries the charge
\begin{equation}\label{eq:Q[kappa]_Lor_taub_nut}
    Q[\kappa] = 8\pi^2 R_y N\kappa_0
\end{equation}
where the $\kappa$-tensor is chosen to be irreducible with non-zero components $\kappa_{t|y} = \kappa_{y|t} \equiv \kappa_0 = \text{constant}$. (More generally, we could take a reducible $\kappa$-tensor with $\kappa_{t|y}  \equiv \kappa_0 = \text{constant}$ and with any $ \kappa_{y|t}$ and get the same charge.) 
The dual is
\begin{equation}\label{eq:tilde_kappa_components}
    \tilde{\kappa}_{abct|t} = \tilde{\kappa}_{abcy|y} = -\epsilon_{abc} \kappa_0
\end{equation}
and all other components of $\tilde{\kappa}$ which are not related to these by anti-symmetry vanish. Indeed, this $\tilde{\kappa}$ satisfies eq.~\eqref{eq:div_ktilde} and is not trivial, as discussed in section~\ref{sec:kappa_cohomology}.

As we have noted, the form of $V$ given in eq.~\eqref{eq:V_kappa} is not necessarily well-defined on spacetimes $\M$ with toroidal directions, such as the present example. 
This example was studied in section~\ref{sec:kappa_subcase 2}, where a   \pqi{4,2}-tensor $V$ was given in eq.~\eqref{eq:second_example_V} which is well-defined on $\M$ and satisfies eq.~\eqref{eq:divV_kappa} for this choice of $\tilde{\kappa}$. We repeat the components of $V$ here for convenience:
\begin{equation}\label{eq:working_1}
    V_{abct|dt} = V_{abcy|dy} = -\frac{1}{8} \kappa_0 \epsilon_{abc} x_d
\end{equation}
All other components not related to these by antisymmetry vanish.

We will evaluate the covariant charge $\mathbb{Q}[\kappa]$ for this choice of $\kappa$ with the integration surface taken to be the codimension-2 cycle $\Sigma_3 = S^2 \times S^1$ where the first factor is a 2-sphere at fixed $r=\abs{x}$ in a constant time slice of $\mathbb{R}^{1,3}$ and the second factor is the circle parameterised by $y$.
Using eq.~\eqref{eq:F=*dX}, we find
\begin{equation}\label{eq:working_2}
    \Omega[V]_{at} = N\kappa_0 \frac{x_a}{\abs{x}^3}
\end{equation}
Then integrating over $\Sigma_3$ gives the charge
\begin{equation}\label{eq:TaubNUT_charge}
    \mathbb{Q}[\kappa] = \int_{\Sigma_3} \star \Omega[V] = 16 \pi^2 R_y N \kappa_0
\end{equation}
We note that this is a factor of two larger than $Q[\kappa]$ in eq.~\eqref{eq:Q[kappa]_Lor_taub_nut}. Indeed, the remainder is contributed by the $\int\dd\star\Phi$ term in eq.~\eqref{eq:Q[kappa]_covariant}, which evaluates to
\begin{equation}\label{eq:TaubNUT_excess}
    \int_{\Sigma_3} \dd\star\Phi = 8\pi^2 R_y N \kappa_0
\end{equation}
as required.

As discussed in the previous example, the fact that the two contributions $Q[\kappa]$ and $\int\dd\star\Phi$ in eqs.~\eqref{eq:Q[kappa]_Lor_taub_nut} and \eqref{eq:TaubNUT_excess} are equal is not a gauge-invariant statement.
Similar considerations about gauge dependence to those below eq.~\eqref{eq:Q=Q_gauge} are true in this case also.
Implementing a gauge transformation $h_{\mu\nu} \to h'_{\mu\nu} = h_{\mu\nu} + \partial_{(\mu} \xi_{\nu)}$ with gauge parameter $\xi_a = 4tA_a$ gives rise to a solution for which $\mathbb{Q}[\kappa] = Q[\kappa]$ and $\int\dd\star\Phi=0$. In this gauge $\Gamma\indices{_{\mu\nu|}^\mu}=0$. In the dual graviton description, the Taub-NUT solution carries an electric charge and the $\int\dd\star\Phi$ contribution vanishes as this measures a magnetic charge. 

\paragraph{Higher-dimensional solutions.}

As in the previous sub-section, we now consider higher-dimensional solutions with a background space $(\mathbb{R}^{1,3} \times T^{d-4})^\bullet$ for $d>5$. As above, the coordinates on $T^{d-4}$ are $y^i$ with $i=1,\dots,d-4$. Then we can again take the Ansatz \eqref{eq:TaubNUT_metric} such that the solution in question is a product of the linearised version of Lorentzian Taub-NUT on $\mathbb{R}^{1,3}$ with a $(d-4)$-torus. Again, in order to generate a non-trivial charge, $\kappa$ is required to be closed but not exact. We take it to have components
\begin{equation}
    \kappa_{ty_2 \dots y_{d-4}|y_1} = \kappa_{y_1 y_2\dots y_{d-4}|t} = \kappa_0
\end{equation}
This involves the volume form on $T^{d-5} \subset T^{d-4}$, so is $\dd_L$-closed and not $\dd_L$-exact. Then evaluating its Hodge dual gives precisely the components in eq.~\eqref{eq:tilde_kappa_components} as before. The manipulations in eqs.~\eqref{eq:working_1} and \eqref{eq:working_2} are unchanged, but the surface of integration must be codimension-2 in the $d$-dimensional background space so we take it to be $\Sigma_{d-2} = S^2 \times T^{d-4} \equiv \Sigma_3 \times T^{d-5}$. Then evaluating the charge on $\Sigma_{d-2}$ yields
\begin{equation}
    \mathbb{Q}[\kappa] = \int_{\Sigma_{d-2}} \star \Omega[V] = 8 \pi N \kappa_0 \; \text{Vol}_{T^{d-4}}
\end{equation}
This clearly reduces to eq.~\eqref{eq:TaubNUT_charge} for $d=5$.

\section{General construction of covariant conserved charges}
\label{sec:general_symmetries}

In section~\ref{sec:kappa_subcase}, we have seen that the 2-forms $\Omega[V]$ given in eq.~\eqref{eq:Omega[V]} include the 2-forms corresponding to both the Penrose and $\kappa$-type charges. We now give a systematic discussion of the constraints that $V$ must satisfy in order for $\Omega[V]$ to be conserved when $R_{\mu\nu}=0$, and examine the charges that can result. We will find that the non-trivial solutions are in correspondence with the three types of secondary currents mentioned previously, $J[k]$, $J[\lambda]$ and $J[\kappa]$, together with  identically conserved topological currents.

A priori, the most general gauge-invariant 2-form linear in the Riemann tensor can be written
\begin{equation}\label{eq:Om}
    \Omega[V]_{\mu\nu} = R^{\alpha\beta\gamma\delta}V_{\mu\nu|\alpha\beta|\gamma\delta}
\end{equation}
for some rank-6 tensor $V$ with the symmetries
\begin{equation}\label{eq:V_222}
    V_{\mu\nu|\alpha\beta|\gamma\delta} = V_{[\mu\nu]|[\alpha\beta]|[\gamma\delta]}
\end{equation}
that is, $V$ is a \pq{2,2,2}-tensor. 
We note that the contraction with the Riemann tensor in eq.~\eqref{eq:Om} projects onto the part of $V$ which is symmetric under the exchange of the $(\alpha\beta)$ and $(\gamma\delta)$ indices. We choose not to make this symmetry explicit in our discussion of $V$ below as it obscures other anti-symmetries of $V$ which we wish to make manifest.
We consider pure gravity in regions where 
\begin{equation} \label{eq:EOM_and_Bianchi}
    R_{\alpha\beta}=0
\end{equation}
and demand the conservation of $\Omega[V]_{\mu\nu}$,
\begin{equation} \label{eq:divOmegaWorking}
    \partial^\mu \Omega[V]_{\mu\nu} = (\partial^\mu R^{\alpha\beta|\gamma\delta}) V_{\mu\nu|\alpha\beta|\gamma\delta} + R^{\alpha\beta|\gamma\delta} \partial^\mu V_{\mu\nu|\alpha\beta|\gamma\delta} = 0
\end{equation}
The first term in eq.~\eqref{eq:divOmegaWorking} will vanish via the differential Bianchi identity $\partial_{[\mu}R_{\alpha\beta]\gamma\delta}=0$ if $V_{\mu\nu|\alpha\beta|\gamma\delta}$ is anti-symmetric in the $\mu$, $\alpha$, and $\beta$ indices.\footnote{Alternatively, $V_{\mu\nu|\alpha\beta|\gamma\delta}$ could be anti-symmetric in the $\mu$, $\gamma$, and $\delta$ indices but, as mentioned above, the contraction with the Riemann tensor imposes that only the part of $V$ that is symmetric under interchange of $(\alpha\beta)$ with $(\gamma\delta)$ contributes to the current, so we choose the former case without loss of generality.} This can be achieved by restricting $V$ to be a \pq{4,2}-tensor, i.e.,
\begin{equation}
    V_{\mu\nu\alpha\beta|\gamma\delta} = V_{[\mu\nu\alpha\beta]|[\gamma\delta]}
\end{equation}
We note that in general $V$ does not have to be an \pq{4,2}-tensor in order for $\Omega[V]$ to be conserved. Indeed, there are other \pq{2,2,2}-tensors which can be used to define a conserved $\Omega[V]$. However, we show in appendix~\ref{app:representation_theory} that the conserved 2-forms $\Omega[V]$ which arise when $V$ is such a tensor have a specific form which can also be accounted for by a different choice of $V$ which is a \pq{4,2}-tensor. Therefore, as our interest lies in classifying the conserved 2-forms $\Omega[V]$ (rather than the tensors $V$ themselves) it is sufficient to restrict $V$ to be a \pq{4,2}-tensor. 

The \pq{4,2} representation which $V$ transforms under is a tensor product of the fully antisymmetric $\four$ and $\two$ representations, so is reducible. We also show in appendix~\ref{app:representation_theory} that the only irreducible representation in the decomposition of the reducible \pq{4,2} representation which contributes non-trivially to $\Omega[V]$ is the one labelled by the Young tableau $\fourtwo$ (i.e. the \textit{irreducible} \pqi{4,2} representation). Therefore, we can take $V$ to be an irreducible \pqi{4,2}-tensor; that is, it must satisfy
\begin{equation}\label{eq:V_irreducible}
    V_{[\mu\nu\alpha\beta|\gamma]\delta}=0
\end{equation}
With these facts in hand, the first term in eq.~\eqref{eq:divOmegaWorking} vanishes. So the requirement that $\Omega[V]$ is conserved reduces to
\begin{equation} \label{eq:divOmegaWorking3} 
    \partial^\mu \Omega[V]_{\mu\nu} = R^{\alpha\beta\gamma\delta} \partial^\mu V_{\mu\nu\alpha\beta|\gamma\delta} = 0
\end{equation}
We now look for constraints on $V$ such that $\partial^\mu \Omega[V]_{\mu\nu}=0$ when eq.~\eqref{eq:EOM_and_Bianchi} is employed (that is, when the field equations $G_{\mu\nu}=T_{\mu\nu}$ are used and sources are absent, $T_{\mu\nu}=0$). We show in Appendix~\ref{app:ansatz} that the most general such constraint is 
\begin{equation}\label{eq:final_V_ansatz}
    \partial^\mu V\indices{_{\mu\nu\alpha\beta}^{|\gamma\delta}} = \omega\indices{_{\nu\alpha\beta}^{[\gamma|\delta]}} + \delta^{[\gamma}_{[\nu} \zeta\indices{_{\alpha\beta]}^{\delta]}} + \delta^{\gamma\delta}_{[\nu\alpha} \theta_{\beta]}
\end{equation}
where $\theta$ is a 1-form, $\zeta$ is a 3-form and $\omega_{\nu\alpha\beta\gamma|\delta}$ is an irreducible \pqi{4,1}-tensor which is traceless, i.e. $\omega\indices{_{\nu\alpha\beta\gamma|}^\gamma}=0$.
This constraint is reached from representation theory considerations alone, and the reasoning is as follows. The left-hand side of eq.~\eqref{eq:final_V_ansatz} is $\dd_L^\dag V$, which is a reducible \pq{3,2}-tensor and so can be written as a sum of $GL(d,\mathbb{R})$ representations. These can be branched into $SO(1,d-1)$ representations, then eq.~\eqref{eq:divOmegaWorking3} demands that several representations which appear in the decomposition cannot contribute. In particular, the constraint~\eqref{eq:final_V_ansatz} is equivalent to the statement that the only irreducible $SO(1,d-1)$ representations which can contribute to $\dd_L^\dag V$ are $\fourone$, $\three$ and $\one$. The tensors $\omega$, $\zeta$, and $\theta$ are given by
\begin{align}
    \omega &= \mathcal{Y}^{SO}_{\fourone}(\dd_L^\dag V) \label{eq:omegabar}\\
    \zeta &= \mathcal{Y}^{SO}_{\three}(\dd_L^\dag V) \\
    \theta &= \mathcal{Y}^{SO}_{\one}(\dd_L^\dag V) \label{eq:theta}
\end{align}
The constraint~\eqref{eq:final_V_ansatz} is then simply the statement that all other irreducible $SO(1,d-1)$ representations in the decomposition of $\dd_L^\dag V$ must vanish. 
In other words, we must demand that
\begin{equation}\label{eq:V_ansatz_reps}
    \mathcal{Y}^{SO}_{\five} (\dd_L^\dag V) = \mathcal{Y}^{SO}_{\threetwo} (\dd_L^\dag V) = \mathcal{Y}^{SO}_{\twoone} (\dd_L^\dag V) = 0
\end{equation}
(Recall that tensors in irreducible $SO(1,d-1)$ representations are traceless so that the Young projector removes   all traces.\footnote{This is the  same method of argument used to determine that $K$ must be a CKY tensor when studying the Penrose 2-form $Y[K]$ in section~\ref{sec:Penrose_review}. In that case, we explicitly wrote the irreducible parts of $\partial K$ in terms of anti-symmetrisations and traces (e.g. in eq.~\eqref{eq:dK_rep_decomp}). With the more complicated representations involved in the case at hand, such formulae are unwieldy and opaque. Instead, we write expressions in terms of $\omega$, $\zeta$ and $\theta$. These objects could be written explicitly in terms of $V$ by evaluating the Young projectors in eqs.~\eqref{eq:omegabar}--\eqref{eq:theta}.}) 
The constraints~\eqref{eq:V_ansatz_reps} on $V$ are analogous to the constraints that $k$ be a Killing vector or that $K$ be a CKY tensor, so can be interpreted as constraining $V$ to be a particular kind of generalised Killing tensor.

Any solution of $\dd_L^\dag V=0$ can be added to a solution of eq.~\eqref{eq:final_V_ansatz} to generate another solution. 
We will first analyse solutions of the homogeneous equation $\dd_L^\dag V=0$ and then analyse particular solutions of eq.~\eqref{eq:final_V_ansatz}. The general solution of \eqref{eq:final_V_ansatz} is then given by a particular solution plus a solution of the homogeneous equation.

\subsection{Solutions of $\dd_L^\dag V=0$}
\label{sec:case0}

We first consider \pqi{4,2}-tensors $V_0$ for which $\dd_L^\dag V_0=0$, that is
\begin{equation}\label{eq:divV=0}
    \partial^\mu (V_0)_{\mu\nu\alpha\beta|\gamma\delta} = 0
\end{equation}
If $V_0$ is $\dd_L$-coexact, then
\begin{equation}\label{eq:V=divVhat}
    (V_0)_{\mu\nu\alpha\beta|\gamma\delta} = \partial^\sigma \hat{V}_{\sigma\mu\nu\alpha\beta|\gamma\delta}
\end{equation}
for some \pq{5,2}-tensor $\hat{V}$, and eq.~\eqref{eq:divV=0} is satisfied. In this case we have 
\begin{equation}
\label{deriveav}
    \Omega[V_0]_{\mu\nu} = -\partial^\alpha \left( R^{\sigma\beta\gamma\delta} \hat{V}_{\sigma\mu\nu\alpha\beta|\gamma\delta} \right)
\end{equation}
where the differential Bianchi identity $\partial^{[\alpha} R^{\sigma\beta]\gamma\delta}=0$ has been used.
That is, $\Omega[V_0]$ is the divergence of a globally defined 3-form (since the 3-form depends only on the curvature) so the charge $\mathbb{Q}[V_0] = \int\star\Omega[V_0]=0$ by Stokes' theorem as the integral is taken over a cycle. 

Therefore, non-trivial charges can only arise from $\dd_L$-coclosed \pqi{4,2}-tensors (satisfying eq.~\eqref{eq:divV=0}) that are not $\dd_L$-coexact (satisfying eq.~\eqref{eq:V=divVhat}), and the charge depends only on $V_0$ modulo a $\dd_L$-coexact piece.
As a result, the charge corresponds to the $\dd_L$-cohomology class of the \pq{d-4,2}-tensor $\star_L V_0$ and this can only be non-zero if the manifold has non-trivial $H^{d-4}$. 

If $V_0$ is $\dd_L$-coclosed then there is a locally-defined (i.e.\ defined in a contractible patch) \pq{5,2}-tensor $\hat{V}$ satisfying \eqref{eq:V=divVhat} so that $\Omega[V_0]$ is a total derivative  
\begin{equation}
\label{deriveavW}
    \Omega[V_0]_{\mu\nu} = \partial^\rho W_{\mu\nu\rho}
\end{equation}
where the 3-form $W$ can be read off from \eqref{deriveav}.
If $V_0$ is not $\dd_L$-coexact, then $\hat V$ will not be a globally-defined gauge-invariant tensor. 
Recalling the discussion of section~\ref{sec:kappa_symms}, if $W$ has non-covariant dependence on the graviton field $h$ then this extra term is not co-exact when $h$ is non-globally defined. Therefore these terms can produce non-trivial charges  for such field configurations. 

Alternatively, via similar manipulations to those in eq.~\eqref{eq:Omega_J[kappa]_link} we find that, for any $V_0$ satisfying eq.~\eqref{eq:divV=0}
\begin{equation}
    \Omega[V_0]_{\mu\nu} = 2 \partial^\alpha \left( \Gamma^{\gamma\delta|\beta} (V_0)_{\mu\nu\alpha\beta|\gamma\delta} \right)~,
\end{equation}
giving an alternative expression for $W$.

Then given any solution $V$ to~\eqref{eq:final_V_ansatz}, a new solution is obtained by taking $V\to V+V_0$ where $V_0$ is any solution of~\eqref{eq:divV=0}.
This changes $\Omega[V]_{\mu\nu}\to \Omega[V+V_0]_{\mu\nu} =\Omega[V]_{\mu\nu} + \partial^\rho W_{\mu\nu\rho}$ so that the current is changed by an identically conserved term $d_L^\dagger W$. This will change the charge by the integral of a total derivative, and this change will be non-trivial if $W$ is not globally-defined.

\subsection{Four dimensions}
\label{sec:4d}

We briefly note that in four dimensions the form of $V$ simplifies as it can then be written
\begin{equation}
    V_{\mu\nu\alpha\beta|\gamma\delta} = \frac{1}{2} \epsilon_{\mu\nu\alpha\beta} v_{\gamma\delta}
\end{equation}
for some 2-form $v$. Now
\begin{equation}
    \Omega[V]_{\mu\nu} = \frac{1}{2} R^{\alpha\beta\gamma\delta} \epsilon_{\mu\nu\alpha\beta} v_{\gamma\delta} = (\star R)_{\mu\nu\gamma\delta} v^{\gamma\delta}
\end{equation}
In regions where $R_{\mu\nu}=0$, the Riemann tensor is equal to the Weyl tensor, $R_{\mu\nu\rho\sigma} = C_{\mu\nu\rho\sigma}$, so
\begin{equation}
    \Omega[V]_{\mu\nu} = (\star C)_{\mu\nu\rho\sigma} v^{\rho\sigma} = C_{\mu\nu\rho\sigma}(\star v)^{\rho\sigma} = R_{\mu\nu\rho\sigma} u^{\rho\sigma}
\end{equation}
where we have used $\star C \star = -C$ and defined $u=\star v$. 
This is of the same form as the Penrose 2-form and is conserved if and only if $u$ is a CKY tensor.
Therefore, in $d=4$, the Penrose 2-form is the most general conserved 2-form current.

Indeed, in four dimensions we have seen in section~\ref{sec:Penrose_review} that all the ADM and dual ADM charges of the graviton are already included in the Penrose charges, and the $\kappa$-invariances are not relevant. In all that follows, we restrict to $d>4$.

\subsection{Relation of covariant and primary currents}
\label{sec:R.V_relation_to_1forms}

For $d>4$, inserting the general constraint \eqref{eq:final_V_ansatz} for the divergence of $V$ into eq.~\eqref{eq:divOmegaWorking3} implies
\begin{align}
\begin{split}
    \partial^\mu \Omega[V]_{\mu\nu} &= R\indices{^{\alpha\beta}_{\gamma\delta}} \left( \omega\indices{_{\nu\alpha\beta}^{[\gamma|\delta]}} + \delta^{[\gamma}_{[\nu} \zeta\indices{_{\alpha\beta]}^{\delta]}} + \delta^{\gamma\delta}_{[\nu\alpha} \theta_{\beta]} \right) \\
    &= R^{\alpha\beta\gamma\delta} \omega_{\nu\alpha\beta\gamma|\delta} - \frac{1}{3} R\indices{^{\alpha\beta\gamma}_{\nu}} \zeta_{\alpha\beta\gamma} - \frac{2}{3} G_{\nu\alpha}\theta^\alpha \label{eq:sources_divOmega}
\end{split}
\end{align}
As guaranteed by the derivation above, this vanishes by the algebraic Bianchi identity $R_{[\alpha\beta\gamma]\delta}=0$ and the field equation $R_{\mu\nu}=0$.

Note that, if we do not impose the Bianchi identity and field equations, the three contributions to eq.~\eqref{eq:sources_divOmega} are of precisely the same form as the 1-form primary currents $j[k]$, $j[\lambda]$ and $j[\kappa]$ defined in eqs.~\eqref{eq:j[k]}, \eqref{eq:j[l]} and \eqref{eq:j[kappa]} respectively. Indeed, if we allow sources for both the graviton and its dual then $R_{\mu\nu}\neq0$ and $R_{[\mu\nu\rho]\sigma}\neq0$ \cite{Hull2001DualityFields}.
The Poincar\'{e} lemma then implies that the covariant 2-form $\Omega[V]$ is related to the secondary currents $J[k]$, $J[\lambda]$ and $J[\kappa]$ up to the divergence of a 3-form locally.
We will see a concrete realisation of this in the following.

\subsection{Sub-cases}

A full solution of the constraint \eqref{eq:final_V_ansatz} for $V$ will be given in section~\ref{sec:general_solution}. In order to illustrate the methods used, we first solve several simpler sub-cases. We begin with the case where only $\zeta$ and $\theta$ are non-zero in section~\ref{sec:Penrose_int_conds}. Next, we consider the case where only $\omega$ is non-zero in section~\ref{sec:kappa_solution}, and finally we include all three terms in the constraint in section~\ref{sec:general_solution}. As discussed above, $V$ is only fixed up to the addition of a solution $V_0$ of the homogeneous equation~\eqref{eq:divV=0}, which adds a derivative term  $\partial^\rho W_{\mu\nu\rho}$ to $\Omega[V]_{\mu\nu}$.

\subsubsection{Case 1: $\omega=0$ --- Penrose charges}
\label{sec:Penrose_int_conds}

Firstly, we study solutions to the constraint \eqref{eq:final_V_ansatz} with $\omega=0$. We will show that, up to the addition of a $\dd_L$-coclosed tensor, the unique solution for $V$ takes the form of the solution presented in eq.~\eqref{eq:V_Penrose}, for which $\Omega[V]=Y[K]$, so that we recover the Penrose charges. 

Given the role that the Penrose solution~\eqref{eq:V_Penrose} will play, we first verify that it solves the constraint \eqref{eq:final_V_ansatz}. This must be the case, since we know $Y[K]$ is conserved.
With $V$ given by eq.~\eqref{eq:V_Penrose}, using the CKY condition~\eqref{eq:CKY_equation}, we find
\begin{align}
    \partial^\mu V\indices{_{\mu\nu\alpha\beta}^{|\gamma\delta}} &= 6\delta^{[\gamma}_{[\mu} \delta^{\delta]}_{\nu} \partial^\mu K_{\alpha\beta]} \nonumber \\
    &= 6\delta^{[\gamma}_{[\mu} \delta^{\delta]}_{\nu} \tilde{K}\indices{_{\alpha\beta]}^\mu} + 12 \delta^{[\gamma}_{[\mu} \delta^{\delta]}_{\nu} \delta^\mu_\alpha \hat{K}_{\beta]} \nonumber \\
    &= -3 \delta^{[\gamma}_{[\nu} \tilde{K}\indices{_{\alpha\beta]}^{\delta]}} +3(d-3) \delta^{[\gamma}_{[\nu} \delta^{\delta]}_\alpha \hat{K}_{\beta]} \label{eq:Penrose_V_working}
\end{align}
So $V$ given in eq.~\eqref{eq:V_Penrose} is indeed a solution of the constraint \eqref{eq:final_V_ansatz} with
\begin{equation} \label{eq:Penrose_theta_zeta}
    \omega_{\nu\alpha\beta\gamma|\delta} = 0 \qc \zeta_{\mu\nu\rho} = -3\tilde{K}_{\mu\nu\rho} \qc \theta_\mu = 3(d-3) \hat{K}_\mu
\end{equation}
From the properties of CKY tensors discussed in section~\ref{sec:Penrose_review}, we see that $\zeta$ is a rank-3 closed CKY tensor and $\theta$ is a Killing vector in this case. 

We now turn to the main goal of this sub-section, to study solutions of the constraint \eqref{eq:final_V_ansatz} with $\omega=0$. That is, we wish to find the \pqi{4,2}-tensors $V$ which satisfy
\begin{equation}\label{eq:divV_omega=0}
    \partial^\mu V\indices{_{\mu\nu\alpha\beta}^{|\gamma\delta}} = \delta^{[\gamma}_{[\nu} \zeta\indices{_{\alpha\beta]}^{\delta]}} + \delta^{[\gamma}_{[\nu} \delta^{\delta]}_{\alpha} \theta_{\beta]}
\end{equation}
In terms of irreducible $SO(1,d-1)$ representations, this is the sub-case of eq.~\eqref{eq:final_V_ansatz} where, as well as the constraints of eq.~\eqref{eq:V_ansatz_reps}, we also have
\begin{equation}
    \mathcal{Y}^{SO}_{\fourone}(\dd_L^\dag V)=0
\end{equation}
We begin by finding integrability conditions on $\zeta$ and $\theta$ which follow from eq.~\eqref{eq:divV_omega=0}.
Taking the divergence of eq.~\eqref{eq:divV_omega=0} gives
\begin{align}\label{eq:step1}
    0 = \partial^\nu \left( \delta^{[\gamma}_{[\nu} \zeta\indices{_{\alpha\beta]}^{\delta]}} + \delta^{[\gamma}_{[\nu} \delta^{\delta]}_{\alpha} \theta_{\beta]} \right)
\end{align}
Expanding the anti-symmetrisations, we have
\begin{equation} \label{eq:integrability_condition}
    0 = \frac{1}{6} \left( \partial_\gamma \zeta_{\alpha\beta\delta} - 3\eta_{\gamma[\alpha} \partial^\nu \zeta_{\beta\delta]\nu} + 2\eta_{\delta[\alpha|} \partial_\gamma \theta_{|\beta]} + \eta_{\gamma[\alpha} \eta_{\beta]\delta} \partial^\nu \theta_\nu \right) - (\gamma\leftrightarrow\delta)
\end{equation}
Now, anti-symmetrising this equation over all the indices immediately gives
\begin{equation} \label{eq:zeta_closed}
    \partial_{[\gamma}\zeta_{\alpha\beta\delta]} = 0
\end{equation}
so $\zeta$ is a closed 3-form. Furthermore, contracting eq.~\eqref{eq:integrability_condition} with $\eta^{\alpha\gamma}$ yields
\begin{equation} \label{eq:trace_integrability_condition}
    0 = (3-d) \partial^\alpha \zeta_{\alpha\beta\delta} + (2-d) \left( \partial_\delta \theta_\beta - \eta_{\delta\beta} \partial^\alpha \theta_\alpha \right)
\end{equation}
and further contracting with $\eta^{\beta\delta}$ gives 
\begin{equation}\label{eq:step2}
    \partial^\alpha \theta_\alpha = 0
\end{equation}
Substituting this into eq.~\eqref{eq:trace_integrability_condition} and symmetrising over $\beta$ and $\delta$ gives
\begin{equation} \label{eq:symmetrised_integrability_condition}
    \partial_{(\delta} \theta_{\beta)} = 0
\end{equation}
so we find that $\theta$ is a Killing vector. Eq.~\eqref{eq:trace_integrability_condition} then implies
\begin{equation}\label{eq:non_local_zeta_theta_link}
    (3-d) \partial^\alpha \zeta_{\alpha\beta\delta} = (2-d) \partial_\beta \theta_\delta
\end{equation}
Substituting this back into eq.~\eqref{eq:integrability_condition} and rearranging yields
\begin{equation}
    \left( \partial_\gamma \zeta_{\alpha\beta\delta} - \frac{3}{d-2} \eta_{\gamma[\alpha} \partial^\nu \zeta_{\beta\delta]\nu} \right) - (\gamma\leftrightarrow\delta) = 0
\end{equation}
Antisymmetrising this equation over the $\beta$, $\gamma$ and $\delta$ indices and using eq.~\eqref{eq:zeta_closed} then finally gives
\begin{equation}\label{eq:zeta_CCKY}
    \partial_\alpha \zeta_{\beta\gamma\delta} - \frac{3}{d-2} \eta_{\alpha[\beta} \partial^\nu \zeta_{\gamma\delta]\nu} = 0
\end{equation}
which is the condition for $\zeta$ to be a rank-3 closed CKY tensor. 
Therefore, it follows from~\eqref{eq:divV_omega=0} that $\theta$  must be a Killing vector and and $\zeta$ must be a 
rank-3 closed CKY tensor. Furthermore, they are related non-locally by eq.~\eqref{eq:non_local_zeta_theta_link}.

From the discussion of section~\ref{sec:Penrose_review}, we know that $\theta$ and $\zeta$ can   be written as the divergence and curl of a CKY 2-form respectively (see Appendix A of Ref.~\cite{PAPER1} for further details). The construction works as follows.
A general Minkowski space Killing vector and rank-3 closed CKY tensor can be parameterised by
\begin{align}
\begin{split}\label{eq:zeta_theta}
    \theta_\mu &= \alpha_\mu + \beta_{\mu\nu} x^\nu \\
    \zeta_{\mu\nu\rho} &= \gamma_{\mu\nu\rho} + \tilde{\beta}_{[\mu\nu}x_{\rho]}
\end{split}
\end{align}
where $\alpha$, $\beta$, $\tilde{\beta}$ and $\gamma$ are constant antisymmetric tensors. The relation between $\zeta$ and $\theta$ in eq.~\eqref{eq:non_local_zeta_theta_link} then implies that
\begin{equation}\label{eq:betatilde_beta}
    \tilde{\beta}_{\mu\nu} = - \frac{3}{d-3} \beta_{\mu\nu}
\end{equation}
Now, introducing a rank-2 CKY tensor $K$ as in eq.~\eqref{eq:CKY_Solution} and setting its components to
\begin{equation}\label{eq:CKY_params}
    \B_\mu = - \frac{2}{d-3} \alpha_\mu \qc \C_{\mu\nu\rho} = -\frac{1}{3} \gamma_{\mu\nu\rho} \qc \D_{\mu\nu} = \frac{1}{3(d-3)} \beta_{\mu\nu},
\end{equation}
we find that $\theta$ and $\zeta$ defined in eq.~\eqref{eq:zeta_theta} are related to the CKY tensor $K$ precisely as in eq.~\eqref{eq:Penrose_theta_zeta}. The constant $\A_{\mu\nu}$ term in the CKY decomposition is left undetermined. Substituting the relations in eq.~\eqref{eq:Penrose_theta_zeta} into~\eqref{eq:divV_omega=0} gives
\begin{equation}
    \partial^\mu V\indices{_{\mu\nu\alpha\beta}^{|\gamma\delta}} = -3 \delta^{[\gamma}_{[\nu} \tilde{K}\indices{_{\alpha\beta]}^{\delta]}} +3(d-3) \delta^{[\gamma}_{[\nu} \delta^{\delta]}_\alpha \hat{K}_{\beta]}
\end{equation}
Now, reversing the manipulations in eq.~\eqref{eq:Penrose_V_working} implies
\begin{equation}\label{eq:divV_ddK}
    \partial^\mu V\indices{_{\mu\nu\alpha\beta|}^{\gamma\delta}} = \partial^\mu \left( 6\delta^{[\gamma}_{[\mu} \delta^{\delta]}_\nu K_{\alpha\beta]} \right)
\end{equation}
such that, up to the addition of a $\dd_L$-coclosed tensor, $V$ is given by the solution presented in eq.~\eqref{eq:V_Penrose} which yields the Penrose charges.

Let us now discuss the charges $\mathbb{Q}[V]$ which result from these solutions of eq.~\eqref{eq:divV_omega=0}.
With $V$ satisfying eq.~\eqref{eq:divV_ddK}, the divergence of $\Omega[V]$ is given by
\begin{equation}\label{eq:divOmega=divY}
    \partial^\mu \Omega[V]_{\mu\nu} = R\indices{^{\alpha\beta}_{\gamma\delta}} \partial^\mu \left( 6 \delta^\gamma_{[\mu} \delta^\delta_{\nu} K_{\alpha\beta]} \right) = \partial^\mu \left( R\indices{^{\alpha\beta}_{\gamma\delta}} \delta^\gamma_{[\mu} \delta^\delta_{\nu} K_{\alpha\beta]} \right) = \partial^\mu Y[K]_{\mu\nu}
\end{equation}
where we have used the differential Bianchi identity $\partial_{[\mu} R_{\alpha\beta]\gamma\delta}=0$ in the second equality and eq.~\eqref{eq:Omega[VPen]=Y[K]} in the final equality.
The Poincaré lemma then implies that $\Omega[V]$ is given locally by the Penrose 2-form up to the divergence of a 3-form, i.e.
\begin{equation}\label{eq:Omega_Penrose_relation}
    \Omega[V]_{\mu\nu} = Y[K]_{\mu\nu} + \partial^\rho W_{\mu\nu\rho}
\end{equation}
Up to the addition of such a term, we see that when $\omega=0$ in the constraint \eqref{eq:final_V_ansatz} \emph{the unique solution for $\Omega[V]$ is given by the Penrose 2-form $Y[K]$}. 

\subsubsection{Case 2: $\zeta=\theta=0$ --- $\kappa$-type charges}
\label{sec:kappa_solution}

We now consider the simplification of the constraint \eqref{eq:final_V_ansatz} in which $\theta=0$ and $\zeta=0$, but $\omega$ is non-zero.
In this case, the constraint becomes 
\begin{equation}\label{eq:kappa_V_ansatz}
    \partial^\mu V_{\mu\nu\alpha\beta|\gamma\delta} = \omega_{\nu\alpha\beta[\gamma|\delta]}
\end{equation}
In terms of irreducible $SO(1,d-1)$ representations, this is the sub-case of eq.~\eqref{eq:final_V_ansatz} where, as well as the constraints of eq.~\eqref{eq:V_ansatz_reps}, $\dd_L^\dag V$ also satisfies
\begin{equation}
    \mathcal{Y}^{SO}_{\three}(\dd_L^\dag V) = \mathcal{Y}^{SO}_{\one}(\dd_L^\dag V) = 0
\end{equation}

A straightforward integrability condition of eq.~\eqref{eq:kappa_V_ansatz} is found by taking a further divergence of eq.~\eqref{eq:kappa_V_ansatz},
\begin{equation}
    \partial^\nu \omega_{\nu\alpha\beta[\gamma|\delta]} = 0
\end{equation}
Antisymmetrising this equation on the $\beta$ and $\delta$ indices then implies
\begin{equation} \label{eq:omega_conserved}
    \partial^\nu \omega_{\nu\alpha\beta\gamma|\delta} = 0
\end{equation}
We note that this is the property obeyed by $\tilde{\kappa}$ in eq.~\eqref{eq:div_ktilde}, which points to a relation between the covariant charges related to $\Omega[V]$ and the $\kappa$-charges. 
Recalling that $\omega$ is also required to be traceless, it satisfies all the same properties required of a $\tilde{\kappa}$ tensor (see section~\ref{sec:kappa_symms}).
We conclude that the tensor $\omega$ appearing in the constraint \eqref{eq:kappa_V_ansatz} must be a $\tilde{\kappa}$-tensor. 

Furthermore, we now show that \emph{all} $\tilde{\kappa}$-tensors can be written as in eq.~\eqref{eq:kappa_V_ansatz} for some choice of $V$ on Minkowski space.
In fact, the relevant choice for $V$ which gives rise to any given $\tilde{\kappa}$ tensor has already been given in eq.~\eqref{eq:V_kappa}, which we repeat here for convenience,
\begin{equation}\label{eq:V[kappa]_repeat}
    V[\tilde{\kappa}]_{\mu\nu\alpha\beta|\gamma\delta} = \frac{(-1)^d}{d-1} \tilde{\kappa}_{\mu\nu\alpha\beta|[\gamma} x_{\delta]}
\end{equation}
For this choice of $V$, it then follows from eq.~\eqref{eq:divV_kappa} that eq.~\eqref{eq:kappa_V_ansatz} is satisfied with 
\begin{equation}
    \omega_{\nu\alpha\beta\gamma|\delta} = \frac{(-1)^d}{d-1} \tilde{\kappa}_{\nu\alpha\beta\gamma|\delta}
\end{equation}
Hence, the tensors $\omega$ appearing in constraint \eqref{eq:kappa_V_ansatz} are in one-to-one correspondence with $\tilde{\kappa}$-tensors, and for each such tensor we can construct a conserved 2-form $\Omega[V]$. 

We recall from the discussion in section~\ref{sec:kappa_subcase 2} that the form of $V$ given in eq.~\eqref{eq:V[kappa]_repeat} is not necessarily well-defined when $\M$ has toroidal directions due to the explicit coordinate dependence of $V$. If $\M$ does have compact directions, it may be that there is no well-defined $V$ which satisfies eq.~\eqref{eq:kappa_V_ansatz} for a given $\tilde{\kappa}$. We have, however, seen examples is section~\ref{sec:kappa_subcase 2} of $\kappa$-tensors on such spacetimes and have constructed \pqi{4,2}-tensors $V$ which satisfy eq.~\eqref{eq:divV_kappa} for these $\kappa$-tensors.

Let us now study the charges $\mathbb{Q}[V]$ associated with the solutions of eq.~\eqref{eq:kappa_V_ansatz}. With $V$ given by eq.~\eqref{eq:V_kappa}, these are the same covariant charges constructed in section~\ref{sec:kappa_subcase}, so $\mathbb{Q}[V[\tilde{\kappa}]] = \mathbb{Q}[\kappa]$. In section~\ref{sec:kappa_needs_cohomology}, we saw that $\kappa$ must represent a non-trivial $\dd_L$-cohomology class if the charge $\mathbb{Q}[\kappa]$ is to be non-zero.

Therefore, when we impose the simplification that $\zeta=0$ and $\theta=0$ in the constraint \eqref{eq:final_V_ansatz}, the non-trivial charges $\mathbb{Q}[\kappa]$ are labelled by $\tilde{\kappa}$ tensors which are $\dd_L$-coclosed but not $\dd_L$-coexact, and so correspond to the $\dd_L$-cohomology of \pq{d-4,1}-tensors.
These charges are related to the charges $Q[\kappa] = \int\star J[\kappa]$ by eq.~\eqref{eq:Q[kappa]_covariant}.

\subsection{The general case}
\label{sec:general_solution}

We now consider the full constraint \eqref{eq:final_V_ansatz} with all components $\omega$, $\zeta$ and $\theta$ non-zero. We determine integrability conditions on these tensors by similar manipulations to those in the sub-cases considered in previous sections. 
In section~\ref{sec:Penrose_int_conds}, we addressed the case where $\omega=0$ and found that integrability conditions impose that $\zeta$ and $\theta$ must be a closed CKY 3-form and a Killing vector respectively. They ultimately appeared in a combination which yielded the Penrose 2-form. 
Including $\omega$ will alter the manipulations somewhat, but the result is similar. We emphasise again that we are aiming to find particular solutions to eq.~\eqref{eq:final_V_ansatz}, to which we can always add a $\dd_L$-coclosed \pqi{4,2}-tensor to give another solution.

We begin as in eq.~\eqref{eq:step1}, by taking the divergence $\partial^\nu$ of the constraint~\eqref{eq:final_V_ansatz}, giving
\begin{equation}\label{eq:modified_step1}
    - \partial^\nu \omega_{\nu\alpha\beta[\gamma|\delta]} = \frac{1}{6} \left( \partial_\gamma \zeta_{\alpha\beta\delta} - 3\eta_{\gamma[\alpha} \partial^\nu \zeta_{\beta\delta]\nu} + 2\eta_{\delta[\alpha|} \partial_\gamma \theta_{|\beta]} + \eta_{\gamma[\alpha} \eta_{\beta]\delta} \partial^\nu \theta_\nu \right) - (\gamma\leftrightarrow\delta)
\end{equation}
Antisymmetrising over all the free indices then implies
\begin{equation}
    \partial_{[\gamma} \zeta_{\alpha\beta\delta]} = -3\partial^\nu \omega_{\nu[\alpha\beta\gamma|\delta]}
\end{equation}
which is in contrast to eq.~\eqref{eq:zeta_closed} when $\omega=0$. 
Contracting eq.~\eqref{eq:modified_step1} with $\eta^{\alpha\gamma}$, the left-hand side vanishes as $\omega$ is traceless. So eq.~\eqref{eq:trace_integrability_condition} is unchanged by the presence of $\omega$. Then the manipulations of eqs.~\eqref{eq:step2} to \eqref{eq:non_local_zeta_theta_link} go through unchanged. In particular, this implies that \emph{$\theta$ is a Killing vector even when $\omega\neq 0$}. 

Following the same manipulations which led to eq.~\eqref{eq:zeta_CCKY} now gives
\begin{equation}\label{eq:zeta_almost_CCKY}
    \partial_\alpha \zeta_{\beta\gamma\delta} - \frac{3}{d-2} \eta_{\alpha[\beta} \partial^\nu \zeta_{\gamma\delta]\nu} = 3\partial^\nu \omega_{\nu\beta\gamma\delta|\alpha}
\end{equation}
which recovers eq.~\eqref{eq:zeta_CCKY} when $\omega=0$. Therefore, the presence of $\omega$ implies that $\zeta$ is no longer a rank-3 closed CKY tensor.

Contracting eq.~\eqref{eq:zeta_almost_CCKY} with $\partial^\beta$ annihilates the right-hand side, giving
\begin{equation}
    \partial_\alpha \partial^\beta \zeta_{\beta\gamma\delta}=0
\end{equation}
Therefore,
\begin{equation}\label{eq:div_zeta_constant}
    \partial^\beta \zeta_{\beta\gamma\delta} = c_{\gamma\delta}
\end{equation}
is a constant 2-form. It then follows that $\zeta$ is at most linear in the coordinates, up to the addition of a co-closed 3-form. The most general such object is then 
\begin{equation}\label{eq:zeta_ansatz_1}
    \zeta_{\beta\gamma\delta} = a_{\beta\gamma\delta} + b_{\beta\gamma\delta|\sigma} x^\sigma + \alpha_{\beta\gamma\delta}
\end{equation}
where $\alpha$ is a co-closed 3-form, $a$ is a constant 3-form and $b$ is a constant \pq{3,1}-tensor with $b\indices{_{\gamma\delta\beta|}^{\beta}} = c_{\gamma\delta}$ to satisfy eq.~\eqref{eq:div_zeta_constant}. The standard way to proceed with such manipulations is to insert eq.~\eqref{eq:zeta_ansatz_1} back into eq.~\eqref{eq:zeta_almost_CCKY} to constrain the form of $b$. However, this is less useful in the present case due to the non-zero right-hand side. Instead, a neater way forward is to decompose $b$ into $SO(1,d-1)$ representations,
\begin{equation}
    b_{\beta\gamma\delta|\sigma} = \bar{b}_{\beta\gamma\delta|\sigma} + \frac{3}{d-2} b'_{[\beta\gamma} \eta_{\delta]\sigma}
\end{equation}
where $b'_{\alpha\beta}=b\indices{_{\alpha\beta\gamma|}^{\gamma}}$ is the trace of $b$, and $\bar{b}$ is traceless. 
Now, eq.~\eqref{eq:zeta_ansatz_1} gives
\begin{equation}
    \zeta_{\beta\gamma\delta} = a_{\beta\gamma\delta} + \frac{3}{d-2} b'_{[\beta\gamma} x_{\delta]} + \bar{b}_{\beta\gamma\delta|\sigma} x^\sigma + \alpha_{\beta\gamma\delta}
\end{equation}
However, using the fact that $\bar{b}$ is traceless, we note that the $\bar{b}$ term can be written as the total divergence of a 4-form,
\begin{equation}
    \bar{b}_{\beta\gamma\delta|\sigma} x^\sigma = \frac{4}{d-2} \partial^\alpha \left( x_{[\alpha} \bar{b}_{\beta\gamma\delta]|\sigma} x^\sigma \right)
\end{equation}
and so this term can be absorbed into a redefinition of $\alpha$. Therefore, the most general form of $\zeta$ is
\begin{equation}\label{eq:zeta_CCKY_Sigma}
    \zeta_{\beta\gamma\delta} = \zeta^0_{\beta\gamma\delta} + \alpha_{\beta\gamma\delta}
\end{equation}
where $\zeta^0_{\beta\gamma\delta} = a_{\beta\gamma\delta} + \frac{3}{d-2} b'_{[\beta\gamma}x_{\delta]}$, which is the general form of a rank-3 closed CKY tensor on flat space. So $\zeta$ has contributions from a closed CKY 3-form and a co-closed 3-form.

Substituting eq.~\eqref{eq:zeta_CCKY_Sigma} into eq.~\eqref{eq:zeta_almost_CCKY} then implies
\begin{equation}
    \partial^\nu \omega_{\nu\beta\gamma\delta|\sigma} = \frac{1}{3} \partial_\sigma \alpha_{\beta\gamma\delta}
\end{equation}
This can be rewritten in the form
\begin{equation}\label{eq:div_pi}
    \partial^\nu \pi_{\nu\beta\gamma\delta|\sigma} = 0
\end{equation}
where 
\begin{equation}
    \pi_{\nu\beta\gamma\delta|\sigma} = \omega_{\nu\beta\gamma\delta|\sigma} - \frac{4}{3} \eta_{\sigma[\nu} \alpha_{\beta\gamma\delta]}
\end{equation}
We see that, in contrast to the results of section~\ref{sec:kappa_solution}, $\omega$ is no longer $\dd_L$-coclosed. Instead, eq.~\eqref{eq:div_pi} suggests that $\pi$ is the correct object to be associated with $\tilde{\kappa}$ in this case. However, we see that $\pi$ is not traceless:
\begin{equation}
    \pi\indices{_{\nu\beta\gamma\sigma|}^\sigma} = \frac{d-3}{3} \alpha_{\nu\beta\gamma}
\end{equation}
As mentioned in section~\ref{sec:kappa_symms}, the current $J[\kappa]$ in eq.~\eqref{eq:J[kappa]} is conserved irrespective of whether $\tilde{\kappa}$ is traceless or not. When $\tilde{\kappa}$ is traceless, this charge corresponds directly to an invariance of the dual graviton. Now, we are forced allow a non-zero trace for $\tilde{\kappa}$ in order to associate it with $\pi$, which is $\dd_L$-coclosed. With this subtlety addressed, we make the association
\begin{equation}\label{eq:pi_kappa}
    \pi_{\nu\beta\gamma\delta|\sigma} = \frac{(-1)^d}{d-1} \tilde{\kappa}_{\nu\beta\gamma\delta|\sigma}
\end{equation}
Then, as in section~\ref{sec:kappa_solution}, a solution for $V$ which is related to this $\pi$ by eq.~\eqref{eq:divV_kappa} is given by $V[\tilde{\kappa}]$ in eq.~\eqref{eq:V[kappa]_repeat}. Again, if $\tilde{\kappa}$ is $\dd_L$-coexact then the associated charge $\mathbb{Q}[\kappa]$ vanishes and non-trivial charges associated with $\tilde{\kappa}$ correspond to the $\dd_L$-cohomology of \pq{d-4,1}-tensors.
As noted in section~\ref{sec:kappa_subcase}, we point out that the $V$ given in eq.~\eqref{eq:V_kappa} is not well-defined when $\M$ has compact dimensions and in that case it is not guaranteed that all $\tilde{\kappa}$ tensors have a corresponding $V[\tilde{\kappa}]$.

Therefore, we find that including all three terms in the constraint \eqref{eq:final_V_ansatz} does not greatly change the conclusions reached in the previous two sections. Namely, $\theta$ must be a Killing vector, $\zeta$ comprises a rank-3 closed CKY tensor and a co-closed 3-form $\alpha$ which combines with $\omega$ in such a way that the \pq{4,1}-tensor $\pi$ is $\dd_L$-coclosed and so must be a $\tilde{\kappa}$ tensor (although now it is not traceless). 

The covariant charges of $\Omega[V]$ associated with these tensors are then related to the secondary currents $J[\kappa]$, $J[\lambda]$ and $J[k]$ as follows. By analogous manipulations to eq.~\eqref{eq:Omega_J[kappa]_link}, up to the addition of a coclosed 2-form, we have
\begin{align}
    \Omega[V]_{\mu\nu} &= -2 \Gamma^{\gamma\delta|\beta} \partial^\alpha V_{\mu\nu\alpha\beta|\gamma\delta} + \partial^\alpha \Phi_{\mu\nu\alpha} \nonumber \\
    &= -2 \Gamma^{\gamma\delta|\beta} \left( \omega_{\mu\nu\beta\gamma|\delta} + \eta_{\gamma[\mu} \zeta_{\nu\beta]\delta} + \eta_{\gamma[\mu} \theta_\beta \eta_{\nu]\delta} \right) + \partial^\alpha \Phi_{\mu\nu\alpha} \nonumber \\
    &= -2 \Gamma^{\gamma\delta|\beta} \left( \omega_{\mu\nu\beta\gamma|\delta} + \eta_{\gamma[\mu} \alpha_{\nu\beta]\delta} + \eta_{\gamma[\mu} \zeta^0_{\nu\beta]\delta} + \eta_{\gamma[\mu} \theta_\beta \eta_{\nu]\delta} \right) + \partial^\alpha \Phi_{\mu\nu\alpha}
\end{align}
where $\Phi$ is the 3-form given in eq.~\eqref{eq:Phi}. We have used the constraint \eqref{eq:final_V_ansatz} in the second equality and eq.~\eqref{eq:zeta_CCKY_Sigma} in the third. Now, writing the Killing vector $\theta$ as the divergence of a CKY 2-form $K$ and writing the closed CKY 3-form $\zeta^0$ as its curl, the final two terms of the previous equation rearrange into the Penrose current as explained in section~\ref{sec:Penrose_int_conds}. 
The first two terms in the previous equation can also be straightforwardly rearranged, yielding
\begin{equation}\label{eq:Omega_decomp_Penrose}
    \Omega[V]_{\mu\nu} = J[\kappa]_{\mu\nu} + Y[K]_{\mu\nu} + \partial^\alpha \Phi_{\mu\nu\alpha}
\end{equation}
where $\zeta^0$ and $\theta$ are given in terms of $K$ by eq.~\eqref{eq:Penrose_theta_zeta} and $\pi$ is related to $\tilde{\kappa}$ by eq.~\eqref{eq:pi_kappa}. Then from eq.~\eqref{eq:Y=J[k]+J[l]+dZ}, we have
\begin{equation}\label{eq:final_Omega_decomp}
    \Omega[V]_{\mu\nu} = J[k]_{\mu\nu} + J[\lambda]_{\mu\nu} + J[\kappa]_{\mu\nu} + \partial^\alpha (\Phi_{\mu\nu\alpha} + Z[K]_{\mu\nu\alpha})
\end{equation}
with $k$ and $\tilde{\lambda}$ given by eqs.~\eqref{eq:k_Khat} and \eqref{eq:lambdatilde_Ktilde} respectively (they are proportional to $\theta$ and $\zeta$). We recall from Refs.~\cite{Benedetti2023GeneralizedGravitons,PAPER1} that the Penrose 2-form $Y[K]$ is co-exact in $d>4$ when $K$ is a KY tensor. So non-trivial charges result only from the $\B$- and $\D$-type CKY tensors in the $Y[K]$ contribution to eq.~\eqref{eq:Omega_decomp_Penrose}.

In summary, in this section we have studied solutions to the general constraint on $V$ in eq.~\eqref{eq:final_V_ansatz}. We have looked for particular solutions to this equation, to which one can always add a \pqi{4,2}-tensor which satisfies eq.~\eqref{eq:divV=0}. Such extra terms can contribute to the charges $\mathbb{Q}[V]$ on spaces with non-trivial cohomology  (see section~\ref{sec:case0}). We then analysed the conserved 2-forms $\Omega[V]$ associated with the solutions of eq.~\eqref{eq:final_V_ansatz}, and their relation to the secondary currents $J[k]$, $J[\lambda]$ and $J[\kappa]$. One can always add a coclosed 2-form to $\Omega[V]$ without spoiling conservation. If this 2-form is locally the divergence of a non-covariant 3-form, it can produce non-trivial charges. Since the addition of such a term is arbitrary, we only classify the charges $\mathbb{Q}[V]$ up to the addition of such terms.

\section{Dimensional reduction of charges in Kaluza-Klein theory}

In the previous section we have given a general discussion of the gauge-invariant 2-form currents which are conserved in the linearised graviton theory in $d$ dimensions. We now consider a toroidal compactification of this theory on $T^n$ and ask whether the conserved quantities in $D\equiv d-n$ dimensions can be related to those in $d$ dimensions.

Concretely, we consider $\mathcal{M}=\mathbb{R}^{1,D-1}\times T^n$ and denote the coordinates $x^\mu = (x^m,y^i)$ where $y^i\sim y^i+2\pi R_i$. The metric on $\mathcal{M}$ is the Minkowski metric $\eta_{\mu\nu}$. 

\subsection{Kaluza-Klein Ansatz}

We will focus in particular on compactifications where the graviton components are independent of the compact directions:
\begin{equation}\label{eq:d_i h=0}
    \partial_i h_{\mu\nu} = 0
\end{equation}
and take an Ansatz of the form
\begin{equation}
    h_{\mu\nu} \quad \longrightarrow \quad h_{mn} = \bar{h}_{mn} - \frac{2}{D-2} \eta_{mn}\phi \qc h_{mi} = 2A_m^{(i)} \qc h_{ij} = 2\phi^{(ij)}
\end{equation}
where $\phi \equiv \sum_{i=1}^n \phi^{(ii)}$. We will refer to $\bar{h}_{mn}$ as the $D$-dimensional graviton. There are also $n$ one-form fields $A^{(i)}_m$ which are the graviphotons and $n(n+1)/2$ scalars $\phi^{(ij)}$ sitting in a symmetric matrix.
Eq.~\eqref{eq:d_i h=0} implies
\begin{equation}\label{eq:dy_R=0}
    \partial_i R_{\mu\nu\rho\sigma} = 0 
\end{equation}
The components of the $d$-dimensional linearised Riemann tensor can be written
\begin{equation}\label{eq:Riemann_components_S1}
\begin{split}
    R_{mnpq} &= \bar{R}_{mnpq} - \frac{2}{D-2} \left( \eta_{m[p}\partial_{q]}\partial_n \phi - \eta_{n[p} \partial_{q]}\partial_m \phi \right) \\ 
    R_{mnpi} &= \partial_p F^{(i)}_{mn} \\ 
    R_{minj} &= \partial_m \partial_n \phi^{(ij)}
\end{split}
\end{equation}
where $\bar{R}_{mnpq} = -2 \partial_{[m} \bar{h}_{n][p,q]}$ is the $D$-dimensional linearised Riemann tensor and $F^{(i)}_{mn} = \partial_m A^{(i)}_n - \partial_n A^{(i)}_m$ is the field strength of the $i^{\text{th}}$ graviphoton.

Here we will consider the theory in the absence of sources, in which case the $d$-dimensional equations of motion imply that the $D$-dimensional fields obey
\begin{equation}\label{eq:EoM_S1}
\begin{split}
    \bar{G}_{mn} &= 0\\
    \partial^p F^{(i)}_{pm} &= 0\\
    \partial_m \partial^m \phi^{(ij)} &= 0
\end{split}
\end{equation}
See Ref.~\cite{PAPER1} for the a discussion of these equations in the presence of sources.

\subsection{2-form currents in the $D$-dimensional theory}

In the dimensionally reduced theory, there are several 2-form currents which can be constructed from the fields $\bar{h}_{mn}$, $A^{(i)}_m$ and $\phi^{(ij)}$. Firstly, we construct the $D$-dimensional 2-forms
\begin{equation}\label{eq:barOmega}
    \bar{\Omega}[\bar{V}]_{mn} = \bar{R}^{abcd} \bar{V}_{mnab|cd}
\end{equation}
where $\bar{V}$ is a \pqi{4,2}-tensor in $D$-dimensions. The considerations of section~\ref{sec:general_symmetries} then apply and $\bar{V}$ must satisfy
\begin{equation}\label{eq:barV_ansatz}
    \partial^m \bar{V}\indices{_{mnab}^{|cd}} = \bar{\omega}\indices{_{nab}^{[c|d]}} + \delta^{[c}_{[n} \zeta\indices{_{ab]}^{d]}} + \delta^{cd}_{[na} \theta_{b]}
\end{equation}
where, $\bar{\omega}$ is a traceless \pqi{4,1}-tensor, $\zeta$ is a 3-form and $\theta$ is a 1-form. Consistency then implies that $\theta$ is a Killing vector, and $\bar{\omega}$ and $\zeta$ assemble into a rank-3 closed CKY tensor and a $\tilde{\kappa}$-tensor as in section~\ref{sec:general_solution}.

In a similar manner, we can construct conserved 2-forms involving the graviphotons. To do so, we contract the $R^{abci}$ component of the $d$-dimensional Riemann tensor \eqref{eq:Riemann_components_S1} with some object. This ensures the result will be gauge-invariant when we uplift to $d$ dimensions. 

There are, in fact, multiple ways to do this but it is sufficient to consider a 2-form of the form
\begin{equation}\label{eq:Delta_def}
    \Delta[\chi]_{mn} = R^{abci} \chi^{(i)}_{mnc|ab} = \partial^c F^{(i)ab} \chi^{(i)}_{mnc|ab}
\end{equation}
for some \pqi{3,2}-tensors $\chi^{(i)}$.
We could also consider a 2-form constructed like $R^{abci}\xi^{(i)}_{mnab|c}$ for some \pqi{4,1}-tensors $\xi^{(i)}$, however we find that these merely produce a subset of the charges of those produced by the $\Delta[\chi]_{mn}$ above.

We must require that $\Delta[\chi]$ is conserved. This will be the case provided that the $\chi^{(i)}$ obey a differential condition. This can be derived analogously to the manipulations in Appendix~\ref{app:ansatz} as follows. $\partial^m \chi^{(i)}_{mnc|ab}$ is, a priori, in the $\two\otimes\two$ representation of $GL(d,\mathbb{R})$. This can be decomposed into irreducible representations of $SO(1,d-1)$. Demanding conservation of $\Delta[\chi]$ then implies that only some of these irreducible representations can appear. In this case, the allowed $SO(1,d-1)$ representations are $\four$ and $\bullet$. Therefore, we demand that $\chi^{(i)}$ satisfies
\begin{equation}\label{eq:chi_ansatz}
    \partial^m \chi^{(i)}_{mnc|ab} = \gamma^{(i)}_{ncab} + (\eta_{a[n} \eta_{c]b} \delta^{(i)} - a\leftrightarrow b)
\end{equation}
where $\gamma^{(i)} = \mathcal{Y}^{SO}_{\four}(\dd_L^\dag \chi^{(i)})$ are 4-forms and $\delta^{(i)} = \mathcal{Y}^{SO}_{\bullet}(\dd_L^\dag \chi^{(i)})$ are 0-forms. Contracting the $c$ and $b$ indices and then taking the divergence of this equation implies $\partial_a \delta^{(i)}=0$, so $\delta^{(i)}$ are constants. Given this, taking a divergence of eq.~\eqref{eq:chi_ansatz} implies that the $\gamma^{(i)}$ are coclosed.

We can also construct conserved 2-forms from the scalar fields $\phi^{(ij)}$. The components of the Riemann tensor involving the scalars is given in the final line of eq.~\eqref{eq:Riemann_components_S1}. The most general conserved 2-form that can be built involving this gauge-invariant combination of derivatives of $\phi^{(ij)}$ is of the form 
\begin{equation}\label{eq:Psi_def}
    \Psi[\mu]_{mn} = \partial^a \partial^b \phi^{(ij)} \mu^{(ij)}_{mna|b}
\end{equation}
where $\mu^{(ij)}$ is a \pqi{3,1}-tensor for each $i,j$. Similar arguments to above show that conservation of this 2-form implies that $\mu^{(ij)}$ must satisfy 
\begin{equation}\label{eq:mu_ansatz}
    \partial^m \mu^{(ij)}_{mna|b} = \nu^{(ij)}_{nab}
\end{equation}
where $\nu^{(ij)}$ is a 3-form for each $i,j$, from which it follows that $\nu^{(ij)}$ is coclosed. 

Therefore, we have conserved 2-forms $\bar{\Omega}[\bar{V}]_{mn}$ in \eqref{eq:barOmega} where $\bar{V}$ satisfies \eqref{eq:barV_ansatz}, $\Delta[\chi]_{mn}$ in \eqref{eq:Delta_def} where $\chi^{(i)}$ satisfies \eqref{eq:chi_ansatz}, and $\Psi[\mu]_{mn}$ in \eqref{eq:Psi_def} where $\mu^{(ij)}$ satisfies \eqref{eq:mu_ansatz}.

\subsection*{Charges in the $D$-dimensional theory}

Let us now consider the charges associated with the conserved 2-forms $\bar{\Omega}[\bar{V}]_{mn}$, $\Sigma[\xi]_{mn}$ and $\Psi[\mu]_{mn}$ constructed above. The construction of $\bar{\Omega}[\bar{V}]$ is exactly analogous to that of $\Omega[V]$ in section~\ref{sec:general_symmetries}, just in $D$ dimensions. Therefore, there are associated charges
\begin{equation}
    \bar{\mathbb{Q}}[\bar{V}] = \int_{\Sigma_{D-2}} \star \bar{\Omega}[\bar{V}]
\end{equation}
where $\Sigma_{D-2}$ is a codimension-2 cycle in $D$-dimensions.
The discussion of the $d$-dimensional charges $\mathbb{Q}[V]$ in section~\ref{sec:general_solution} carries through analogously here. 

Next we consider the charges constructed from the conserved 2-forms $\Delta[\chi]$. From eq.~\eqref{eq:chi_ansatz}, we have
\begin{align}
\begin{split}
    \Delta[\chi]_{mn} &= \partial^c (F^{(i)ab} \chi^{(i)}_{mnc|ab}) - F^{(i)ab} ( \gamma^{(i)}_{mnab} + 2\eta_{a[m}\eta_{n]b} \delta^{(i)}) \\
    &= \partial^c (F^{(i)ab} \chi^{(i)}_{mnc|ab}) - F^{(i)ab} \gamma^{(i)}_{mnab} - \delta^{(i)} F^{(i)}_{mn} 
\end{split}
\end{align}
The first term is the divergence of a globally defined 3-form and so will vanish by Stokes' theorem when integrated over a cycle. Therefore, the related charge is
\begin{equation}
    \bar{\mathbb{Q}}[\chi] \equiv \int_{\Sigma_{D-2}} \star \Delta[\chi] = -2 \int_{\Sigma_{D-2}} \star \gamma^{(i)} \wedge F^{(i)} - \delta^{(i)} \int_{\Sigma_{D-2}} \star F^{(i)}
\end{equation}
The first term on the right-hand side is a magnetic charge which depends only on the cohomology class of the closed $(D-4)$-form $\star\gamma^{(i)}$. It is carried by $(D-4)$-branes wrapping cycles which are Poincaré dual to the cohomology class $[\star\gamma^{(i)}]$ \cite{HullYetAppear}. The second term on the right hand side is the electric charge of the graviphotons, which generates the standard electric $U(1)$ 1-form symmetry of Maxwell theory \cite{Gaiotto2015GeneralizedSymmetries}.

Finally we consider the charges which are constructed from the 2-forms $\Psi[\mu]$, related to the scalars $\phi^{(ij)}$. The charge is
\begin{equation}
    \bar{\mathbb{Q}}[\mu] = \int_{\Sigma_{D-2}} \star \Psi[\mu]
\end{equation}
From the definition \eqref{eq:Psi_def}, stripping off derivatives and using eq.~\eqref{eq:mu_ansatz} yields
\begin{equation}
    \Psi[\mu]_{mn} = \partial^a \left( \partial^b \phi^{(ij)} \mu^{(ij)}_{mna|b} - \phi^{(ij)} \nu^{(ij)}_{mna} \right)
\end{equation}
The first term inside the parentheses is globally defined since the gauge transformations of the scalars $\phi^{(ij)}$ are simply shifts by a constant. Therefore, this term does not contribute to the charge. The charge can then be written
\begin{equation}
    \bar{\mathbb{Q}}[\mu] = -\int_{\Sigma_{D-2}} \dd\star(\phi^{(ij)}\nu^{(ij)})
\end{equation}
which is a topological charge and can be non-zero only when $\phi^{(ij)}$ are not globally defined. Physically, this measures the winding of the scalars around the compact directions.

\subsection{Embedding in $d$-dimensional theory}

Having constructed the $D$-dimensional charges which can be constructed from the fields $\bar{h}$, $A^{(i)}$, and $\phi^{(ij)}$, we now wish to embed them into the $d$-dimensional charges $\Omega[V]$ which can be constructed from the $d$-dimensional graviton $h$. 
Firstly, we define a \pqi{4,2}-tensor $V$ in $d$-dimensions with components
\begin{equation}
    V_{mnab|cd} = \bar{V}_{mnab|cd}\qc V_{mnai|cd} = V_{mncd|ai} = \frac{1}{4} \chi^{(i)}_{mna|cd}\qc V_{mnai|bj} = \frac{1}{4}\mu^{(ij)}_{mna|b}
\end{equation}
and a $d$-dimensional \pqi{2,2}-tensor $\mathscr{P}_{\mu\nu\rho\sigma}$ with components
\begin{equation}
    \mathscr{P}_{mnpq} = \frac{2}{D-2} \left( \eta_{m[p} \partial_{q]}\partial_n \phi - m\leftrightarrow n \right)
\end{equation}
and all other components vanishing. Now consider the object
\begin{equation}
    \Omega[V]_{mn} = \bar{\Omega}[\bar{V}]_{mn} + \Delta[\chi]_{mn} + \Psi[\mu]_{mn}
\end{equation}
This is the $mn$ component of the tensor
\begin{equation}
    \Omega[V]_{\mu\nu} = (R^{\alpha\beta\gamma\delta} + \mathscr{P}^{\alpha\beta\gamma\delta})V_{\mu\nu\alpha\beta|\gamma\delta}
\end{equation}
This can be seen by expanding
\begin{align}
    \Omega[V]_{mn} &= (R^{abcd} + \mathscr{P}^{abcd})V_{mnab|cd} + 4 R^{abci} V_{mnab|ci} + 4R^{aibj} V_{mnai|bj}
\end{align}
and substituting in eqs.~\eqref{eq:Riemann_components_S1} and comparing with the definitions \eqref{eq:barOmega}, \eqref{eq:Delta_def} and \eqref{eq:Psi_def}.

Therefore, if we choose to integrate $\star\Omega[V]$ over a $(d-2)$-cycle given by
\begin{equation}
    \Sigma_{d-2} = \Sigma_{D-2} \times T^n
\end{equation}
then
\begin{equation}
    \mathbb{Q}[V] = \int_{\Sigma_{d-2}} \star \Omega[V] = \text{Vol}_{T^n} \left( \bar{\mathbb{Q}}[\bar{V}] + \bar{\mathbb{Q}}[\chi] + \bar{\mathbb{Q}}[\mu] \right)
\end{equation}
Therefore, we see that the gauge-invariant charges $\bar{\mathbb{Q}}[\bar{V}]$, $\bar{\mathbb{Q}}[\chi]$, and $\bar{\mathbb{Q}}[\mu]$ which are constructed in the dimensionally reduced $D$-dimensional theory have a unified gauge-invariant origin in the original $d$-dimensional theory.

\section{Discussion \& outlook}

In this work we have focused on the construction of gauge-invariant 2-form conserved currents in the theory of the linearised graviton on a background spacetime with a flat metric but which may have non-trivial topology due to toroidal compactification or the removal of some regions.
These currents can then be used to define charges which are codimension-2 topological operators. In the quantum theory, these generate 1-form symmetries.
In particular, we have found a covariant expression for the gravitational magnetic charge of \cite{HullYetAppear} constructed from a $[d-4,1]$ Killing tensor $\kappa$ which satisfies the differential constraint \eqref{eq:dkappa=0}.
This charge is interesting as it is carried by the linearised Kaluza-Klein monopole solution. 
 
Furthermore, we have classified all such 2-form currents of the general form \eqref{eq:Omega[V]_intro}. We find that, besides the $\kappa$-currents above, the other non-trivial conserved 2-forms are the Penrose currents \eqref{eq:Penrose2form} which have been discussed previously in \cite{Penrose1982Quasi-localRelativity,Hinterbichler2023GravitySymmetries,Benedetti2023GeneralizedGravitons,PAPER1}.

Our discussion here is purely for the linearised theory. It is of course important and interesting to consider the extension of our results to the non-linear theory of general relativity or supergravity. Kaluza-Klein monopoles play a key role in the non-linear theory and carry the Kaluza-Klein monopole charge of \cite{Hull:braneCharges}. Extending our work to the non-linear case should shed some light onto the significance of this and related charges.
We will discuss this and related issues elsewhere.

There is a general expectation that generalised symmetries should come in dual pairs \cite{CasiniCompleteness, Benedetti2023GeneralizedGravitons}. That is, there should be the same number of continuous $p$-form symmetries as $(d-p-1)$-form symmetries. This was verified for the free graviton on $d$-dimensional Minkowski space in \cite{Benedetti2023GeneralizedGravitons} by showing that some of the Penrose charges are trivial in $d>4$ and showing that there is the same number of conserved $(d-2)$-form currents as of non-trivial Penrose charges. We have shown here that, on topologically non-trivial backgrounds, there are more 1-form symmetries than those generated by the Penrose charges. It would, therefore, be interesting to understand the full set of generalised symmetries that are dual to these on such backgrounds and to see whether they come in dual pairs; we will return to this elsewhere.

\paragraph{Acknowledgements.}
CH is supported by the STFC Consolidated Grants ST/T000791/1 and ST/X000575/1.
MLH is supported by a President's Scholarship from Imperial College London.
UL gratefully acknowledges a Leverhulme visiting professorship to Imperial College as well as the hospitality of the theory group at Imperial.

\appendix

\section{Some definitions and conventions}
\label{app:conventions}

The secondary current $J[k]$ introduced in eq.~\eqref{eq:ADM_charges} can be written
\begin{equation} \label{eq:J[k]_def}
    J[k]^{\mu\nu} = \partial_\sigma \mathcal{K}^{\mu\nu|\rho\sigma} k_\rho - \mathcal{K}^{\mu\sigma|\rho\nu} \partial_\sigma k_\rho
\end{equation}
where
\begin{equation} \label{eq:DefK}
    \mathcal{K}^{\mu\nu|\rho\sigma} = -3\eta^{\mu\nu\alpha|\rho\sigma\beta} h_{\alpha\beta}
\end{equation}
and
\begin{equation}
    \eta^{\mu\nu\rho|\alpha\beta\gamma} = \eta^{\alpha\sigma} \eta^{\beta\kappa} \eta^{\gamma\lambda} \delta^\mu_{[\sigma} \delta^\nu_\kappa \delta^\rho_{\lambda]}
\end{equation}

The secondary current $J[\lambda]$ in eq.~\eqref{eq:j[l]}, corresponding to the $\lambda$-invariances, can be explicitly constructed in terms of the dual graviton $D$ and, when $\tilde{\lambda}$ is constant, can be rewritten in terms of the graviton $h_{\mu\nu}$. In this case, the secondary current is
\begin{equation}\label{eq:J[lambda]=dZ[lambda]}
    J[\lambda]_{\mu\nu} = \partial^\rho Z[\lambda]_{\mu\nu\rho}
\end{equation}
where
\begin{equation}\label{eq:Z[lambda]}
    Z[\lambda]_{\mu\nu\rho} = (-1)^{d+1} \frac{4}{d-1} \tilde \lambda_{[\sigma \mu\nu }h_{\rho]}{}^ \sigma   
\end{equation}
When $\tilde{\lambda}$ is not constant, there is another contribution to $Z[\lambda]$ in the dual theory which cannot be written locally in terms of the graviton $h_{\mu\nu}$. In any case, the current $J[\lambda]$ in eq.~\eqref{eq:J[lambda]=dZ[lambda]} is clearly conserved for any 3-form $\tilde{\lambda}$ so, in the graviton theory, we will take this as the definition of $J[\lambda]$ for all $\tilde{\lambda}$.

In $d=4$, the $\lambda$ tensors reduce to Killing vectors (since a rank-1 KY tensor is simply a Killing vector). The secondary current $\tilde{J}[\tilde{k}]$ which appears in eq.~\eqref{eq:4d_Penrose}, associated to the dual graviton's invariances in four dimensions, can be written
\begin{equation}
    \tilde{J}[\tilde{k}]_{\mu\nu} = \partial^\rho Z[\tilde{k}]_{\mu\nu\rho}
\end{equation}
where
\begin{equation}\label{eq:Z[k]}
    Z[\tilde{k}]_{\mu\nu\rho} = - \epsilon_{\mu\nu\rho\sigma} \tilde{k}^\tau h\indices{_\tau^\sigma}
\end{equation}

\section{Decomposition of $V$ into $\text{GL}(d,\mathbb{R})$ representations}
\label{app:representation_theory}

As discussed in section~\ref{sec:general_symmetries}, the most general form that the tensor $V$ in eq.~\eqref{eq:Omega[V]} can take is that of a reducible \pq{2,2,2}-tensor, satisfying eq.~\eqref{eq:V_222}. That is, it transforms in a $GL(d,\mathbb{R})$ representation characterised by a tensor product Young tableau
\begin{equation}\label{eq:product_tableaux}
    \ydiagram{1,1} \otimes \ydiagram{1,1} \otimes \ydiagram{1,1} 
\end{equation}
We wish to impose constraints on $V$ such that $\partial^\mu \Omega[V]_{\mu\nu}=0$. From eq.~\eqref{eq:divOmegaWorking}, we have
\begin{equation}\label{eq:divOmega_app}
    \partial^\mu \Omega[V]_{\mu\nu} = (\partial^\mu R^{\alpha\beta|\gamma\delta}) V_{\mu\nu|\alpha\beta|\gamma\delta} + R^{\alpha\beta|\gamma\delta} \partial^\mu V_{\mu\nu|\alpha\beta|\gamma\delta} = 0
\end{equation}
This must vanish for all field configurations satisfying the field equation $R_{\mu\nu}=0$ and the Bianchi identities
\begin{equation}\label{eq:bianchis}
    R_{[\mu\nu\rho]\sigma}=0\qc \partial_{[\alpha} R_{\beta\gamma]\rho\sigma}=0\qc \partial^\mu R_{\mu\nu\rho\sigma}=0
\end{equation}
The last of these equations follows from the second by using $R_{\mu\nu}=0$. 
The fact that this must vanish for all such configurations implies that the two terms in eq.~\eqref{eq:divOmega_app} must vanish independently.
The second term in eq.~\eqref{eq:divOmega_app} will vanish provided $V$ satisfies some differential condition which will be determined later (the details of this calculation can be found in Appendix~\ref{app:ansatz}). 
The vanishing of the first term, however, places a constraint on the symmetries of $V$ (i.e. the representation which $V$ transforms in) such that it projects $\partial^\mu R^{\alpha\beta|\gamma\delta}$ onto the field equations and/or Bianchi identities in such a way that the result vanishes. We determine this constraint in this Appendix, proceeding in three stages as follows.

\begin{enumerate}
    \item Firstly, since we are only interested in tensors $V$ for which $\Omega[V]$ is non-zero, we will decompose the na\"{i}ve tensor product representation \eqref{eq:product_tableaux} which $V$ transforms in into irreducible $GL(d,\mathbb{R}$) representations. Then, by contracting with the Riemann tensor to form $\Omega[V]_{\mu\nu} = R^{\alpha\beta\gamma\delta}V_{\mu\nu\alpha\beta|\gamma\delta}$ we will determine which of these can contribute to $V$ in a way which gives a non-trivial $\Omega[V]$ from which charges can be constructed. 
    \item Having found the allowed irreducible $GL(d,\mathbb{R})$ representations which $V$ can transform in, we will insert these into eq.~\eqref{eq:divOmega_app} and demand that the first term vanishes (the second term will be considered in Appendix~\ref{app:ansatz}). Some of the representations which $V$ can transform in immediately project $\partial^\mu R^{\alpha\beta|\gamma\delta}$ onto a Bianchi identity, and so vanish. For example, this is the case if $V$ transforms in the irreducible [4,2] representation, then $\partial^\mu R^{\alpha\beta|\gamma\delta} V_{\mu\nu\alpha\beta|\gamma\delta} = \partial^{[\mu}R^{\alpha\beta]|\gamma\delta}V_{\mu\nu\alpha\beta|\gamma\delta}=0$ from the second Bianchi identity in eq.~\eqref{eq:bianchis}. However, the first term of eq.~\eqref{eq:divOmega_app} could also vanish by the third identity in eq.~\eqref{eq:bianchis} if $V$ has a factor of the Minkowski metric, i.e. is a sum of terms each of which is the tensor product of the Minkowski metric with another tensor. For example, if $V_{\mu\nu|\alpha\beta|\gamma\delta} = \eta_{\alpha[\mu}A_{\nu]|\beta|\gamma\delta} - (\alpha\leftrightarrow\beta)$ for some \pq{1,1,2}-tensor $A$, then $\partial^\mu R^{\alpha\beta|\gamma\delta} V_{\mu\nu|\alpha\beta|\gamma\delta} = 2\partial_\alpha R^{\alpha\beta|\gamma\delta} A_{\nu|\beta|\gamma\delta} = 0$. 
    
    A systematic method to analyse all the ways in which the Minkowski metric can arise in $V$ is to decompose the irreducible $GL(d,\mathbb{R})$ representations into irreducible $SO(1,d-1)$ representations by separating out all possible traces within $V$. We can then find  all the $SO(1,d-1)$ representations that can contribute to $V$ in a manner such that the first term in eq.~\eqref{eq:divOmega_app} vanishes. Indeed, this is precisely the same logic employed in section~\ref{sec:Penrose_review} to show that $K$ must be a CKY in order for the Penrose 2-form $Y[K]$ to be conserved.
    \item Since we are ultimately interested in the set of conserved 2-forms $\Omega[V]$, rather than the tensors $V$ themselves, we will ask whether all the allowed representations which $V$ can transform in produce distinct 2-forms $\Omega[V]$. We will see that, while there are other representations which are allowed in the decomposition of $V$, the form of $\Omega[V]$ which arises can always be constructed with a $V$ in the [4,2] $GL(d,\mathbb{R})$ representation. Therefore, it will be sufficient to simply demand that $V$ is a [4,2]-tensor.
\end{enumerate}

\paragraph{$GL(d,\mathbb{R})$ decomposition of $V$.}

We begin with the first stage outlined above.
Decomposing the tensor product representation in eq.~\eqref{eq:product_tableaux} into irreducible $GL(d,\mathbb{R})$ representations gives 
\begin{equation}\label{eq:2x2x2}
    \two \otimes \two \otimes \two = \left( \; \four \oplus \threeone \oplus \twotwo \;\right) \otimes \two = \left( \;\four\otimes\two\; \right) \oplus \left( \;\threeone \otimes \two \;\right) \oplus \left(\; \twotwo\otimes\two \;\right)
\end{equation}
The three tensor product representations on the right-hand side of eq.~\eqref{eq:2x2x2} can themselves be decomposed similarly. The first of them gives
\begin{align}\label{eq:4x2}
    \ydiagram{1,1,1,1} \otimes \ydiagram{1,1} &= \ydiagram{1,1,1,1,1,1} \oplus \ydiagram{2,1,1,1,1} \oplus \ydiagram{2,2,1,1}
\end{align}
However, we note that only the final diagram gives a non-zero contribution to $\Omega[V]$, because of the Bianchi identity $R_{[\alpha\beta\gamma]\delta}=0$. This can be seen explicitly by decomposing the component of $V$ in the $\four\otimes\two$ representation as
\begin{equation}
    V^{\four\otimes\two}_{\mu\nu\alpha\beta|\gamma\delta} = V^{\six}_{\mu\nu\alpha\beta\gamma\delta} + V^{\fiveone}_{\mu\nu\alpha\beta[\gamma|\delta]} + V^{\fourtwo}_{\mu\nu\alpha\beta|\gamma\delta}
\end{equation}
where the three terms on the right-hand side are the Young projections of $V^{\four\otimes\two}$ onto the representation in the superscript, e.g. $V^{\fourtwo} = \mathcal{Y}^{GL}_{\fourtwo}(V^{\four\otimes\two})$. Contraction with the Riemann tensor then immediately gives
\begin{equation}\label{eq:Omega[4,2]}
    \Omega[V^{\four\otimes\two}]_{\mu\nu} = R^{\alpha\beta\gamma\delta} ( V^{\six}_{\mu\nu\alpha\beta\gamma\delta} + V^{\fiveone}_{\mu\nu\alpha\beta\gamma|\delta} + V^{\fourtwo}_{\mu\nu\alpha\beta|\gamma\delta} ) = R^{\alpha\beta\gamma\delta} V^{\fourtwo}_{\mu\nu\alpha\beta|\gamma\delta}
\end{equation}
where the first two terms in the parentheses vanish upon contraction with the Riemann tensor from $R_{[\alpha\beta\gamma]\delta}=0$.

The second of the three tensor products on the right-hand side of eq.~\eqref{eq:2x2x2} decomposes as
\begin{equation}\label{eq:3,1x2}
    \ydiagram{2,1,1} \otimes \ydiagram{1,1} = \ydiagram{2,1,1,1,1} \oplus \ydiagram{2,2,1,1} \oplus \ydiagram{2,2,2} \oplus \ydiagram{3,2,1} \oplus \ydiagram{3,1,1,1}
\end{equation}
In a similar manner to eq.~\eqref{eq:Omega[4,2]}, some of the representations on the right-hand side of eq.~\eqref{eq:3,1x2} give $\Omega[V]=0$ identically from the Bianchi identity. In particular, we can write the decomposition \eqref{eq:3,1x2} in terms of the tensors $V$ as
\begin{equation}\label{eq:V_3,1x2}
    V^{\threeone\otimes\two}_{\mu\nu\alpha|\beta|\gamma\delta} = U^{\fiveone}_{\mu\nu\alpha\gamma\delta|\beta} + U^{\fourtwo}_{\mu\nu\alpha[\gamma|\delta]\beta} + U^{\threethree}_{\mu\nu\alpha|\beta\gamma\delta} + U^{\threetwoone}_{\mu\nu\alpha|\beta[\gamma|\delta]} + \bigg(\frac{1}{2}U^{\fouroneone}_{\mu\nu\alpha\gamma|\beta|\delta} - (\gamma\leftrightarrow\delta)\bigg)
\end{equation}
where, as before, each term on the right-hand side is the Young projection of $V^{\threeone\otimes\two}$ onto the irreducible $GL(d,\mathbb{R})$ representation indicated in the superscript. Contracting with $R^{\alpha\beta\gamma\delta}$ to form $\Omega[V]_{\mu\nu}$, we immediately see that the $\fiveone$ and $\threethree$ representations give a vanishing result from the Bianchi identity $R_{[\alpha\beta\gamma]\delta}=0$. Furthermore, the $\fouroneone$ representation yields a vanishing result because
\begin{equation}
    R^{\alpha\beta\gamma\delta} V^{\fouroneone}_{\mu\nu\alpha\gamma|\beta|\delta} = \frac{1}{2}R^{\alpha\gamma\beta\delta}V^{\fouroneone}_{\mu\nu\alpha\gamma|[\beta|\delta]} = 0
\end{equation}
where we have used $R_{[\alpha\beta\gamma]\delta}=0$ in the first equality and $V^{\fouroneone}_{\mu\nu\alpha\gamma|[\beta|\delta]}=0$ in the second, which is a criterion for the irreducibility of this representation. Therefore, only the $\fourtwo$ and $\threetwoone$ representations in eq.~\eqref{eq:V_3,1x2} give non-vanishing $\Omega[V]$.

Finally, the third tensor product on the right-hand side of eq.~\eqref{eq:2x2x2} decomposes as
\begin{equation}\label{eq:2,2x2}
    \ydiagram{2,2} \otimes \ydiagram{1,1} = \ydiagram{2,2,1,1} \oplus \ydiagram{3,2,1} \oplus \ydiagram{3,3}
\end{equation}
Again, in terms of the tensor $V$, this decomposition can be written
\begin{equation}
    V^{\twotwo\otimes\two}_{\mu\nu|\alpha\beta|\gamma\delta} = W^{\fourtwo}_{\mu\nu\gamma\delta|\alpha\beta} + \bigg(\frac{1}{2} W^{\threetwoone}_{\mu\nu\gamma|\alpha\beta|\delta} - (\gamma\leftrightarrow\delta) \bigg) + W^{\twotwotwo}_{\mu\nu|\alpha\beta|\gamma\delta}
\end{equation}
where each term on the right-hand side is the Young projection of $V^{\twotwo\otimes\two}$ onto the irreducible $GL(d,\mathbb{R})$ representation labelled.
All of the terms on the right-hand side can contribute to non-zero $\Omega[V]$.

Combining the results from the three tensor products on the right-hand side of eq.~\eqref{eq:2x2x2}, we can write the most general $V$ which contributes to non-zero $\Omega[V]$ as
\begin{align}
\begin{split}\label{eq:V_2x2x2}
    V_{\mu\nu|\alpha\beta|\gamma\delta} &= V^{\fourtwo}_{\mu\nu\alpha\beta|\gamma\delta} + U^{\fourtwo}_{\mu\nu\alpha[\gamma|\delta]\beta} + U^{\threetwoone}_{\mu\nu[\alpha|\beta][\gamma|\delta]} \\
    &\quad + W^{\fourtwo}_{\mu\nu\gamma\delta|\alpha\beta} + \bigg(\frac{1}{2}W^{\threetwoone}_{\mu\nu\gamma|\alpha\beta|\delta} - (\gamma\leftrightarrow\delta) \bigg) + W^{\twotwotwo}_{\mu\nu|\alpha\beta|\gamma\delta} 
\end{split}
\end{align}
Note that there are three tensors in the $\fourtwo$ representation with different index placements. This reflects the fact that this representation arises with multiplicity three in the full decomposition of $\two\otimes\two\otimes\two$ in eq.~\eqref{eq:2x2x2}; that is, it appears once in each of the three tensor products on the right-hand side of eq.~\eqref{eq:2x2x2} (see eqs.~\eqref{eq:4x2}, \eqref{eq:3,1x2} and \eqref{eq:2,2x2}). Similarly, the $\threetwoone$ representation appears twice.

However, while these representations appear with multiple different index placements in the decomposition of $V$, the properties of the Riemann tensor imply that they all contribute in the same way to $\Omega[V]$. Indeed, contracting eq.~\eqref{eq:V_2x2x2} with $R^{\alpha\beta\gamma\delta}$ and using the identities $R_{[\alpha\beta\gamma]\delta}=0$ and $R_{\alpha\beta\gamma\delta} = R_{\gamma\delta\alpha\beta}$ we find
\begin{equation}
\begin{split}
    \Omega[V]_{\mu\nu} &= R^{\alpha\beta\gamma\delta} \bigg( V^{\fourtwo}_{\mu\nu\alpha\beta|\gamma\delta} - \frac{1}{2} U^{\fourtwo}_{\mu\nu\alpha\beta|\gamma\delta} + W^{\fourtwo}_{\mu\nu\alpha\beta|\gamma\delta} \\
    &\qquad\qquad - \frac{1}{2} U^{\threetwoone}_{\mu\nu\alpha|\beta\gamma|\delta} + W^{\threetwoone}_{\mu\nu\alpha|\beta\gamma|\delta} + W^{\twotwotwo}_{\mu\nu|\alpha\beta|\gamma\delta} \bigg)
\end{split}
\end{equation}
This demonstrates that all three terms in the $\fourtwo$ representation contribute to $\Omega[V]$ in the same way. Similarly, both the $\threetwoone$ terms contribute in the same way. Therefore, since we are interested in classifying the possible conserved 2-forms $\Omega[V]$, we can simply choose one of the index permutations for the $\fourtwo$ and $\threetwoone$ representations to include in $V$. This will still produce all the possible 2-forms $\Omega[V]$. Therefore, without loss of generality, we are free to set
\begin{equation}
    V_{\mu\nu|\alpha\beta|\gamma\delta} = V^{\fourtwo}_{\mu\nu\alpha\beta|\gamma\delta} + U^{\threetwoone}_{\mu\nu[\alpha|\beta][\gamma|\delta]} + W^{\twotwotwo}_{\mu\nu|\alpha\beta|\gamma\delta}
\end{equation}

\paragraph{Vanishing of first term of eq.~\eqref{eq:divOmega_app}.}

We now find the constraints on $V$ in order for the first term in eq.~\eqref{eq:divOmega_app} to vanish, as required for $\Omega[V]$ to be conserved. We have
\begin{equation}\label{eq:dRV_decomp}
    \partial^\mu R^{\alpha\beta\gamma\delta} V_{\mu\nu|\alpha\beta|\gamma\delta} = \partial^\mu R^{\alpha\beta\gamma\delta} (V^{\fourtwo}_{\mu\nu\alpha\beta|\gamma\delta} + U^{\threetwoone}_{\mu\nu\alpha|\beta\gamma|\delta} + W^{\twotwotwo}_{\mu\nu|\alpha\beta|\gamma\delta})
\end{equation}
The $\fourtwo$ contribution projects onto the differential Bianchi identity $\partial^{[\mu}R^{\alpha\beta]\gamma\delta}=0$ and so vanishes. The other two terms, however, do not vanish in this manner. Instead, we should make use of the field equation $R_{\mu\nu}=0$, as well as the Bianchi identities in eq.~\eqref{eq:bianchis}. In order to allow for the use of the field equation, we view the irreducible $GL(d,\mathbb{R})$ representations as reducible $SO(1,d-1)$ representations. They are reducible because $SO(1,d-1)$ has the additional invariant tensor $\eta_{\mu\nu}$. 
We can decompose the $\threetwoone$ and $\twotwotwo$ $GL(d,\mathbb{R}$) representations into $SO(1,d-1)$ representations by removing all possible traces. This introduces various factors of the $SO(1,d-1)$-invariant metric $\eta_{\mu\nu}$ which can contract with the Riemann tensor to give the Ricci tensor.

We begin with the $\threetwoone$ representation, we have a decomposition 
\begin{equation}
\begin{split}\label{eq:3,2,1_SO}
    \threetwoone_{GL} &= \left( \; \threetwoone \oplus \; \twooneone \oplus \; \threeone \oplus 3\; \oneone \oplus 2\; \two \oplus 3 \; \bullet \right)_{SO}
\end{split}
\end{equation}
where $\bullet$ represents the trivial $\mathbf{1}$ representation. The degeneracies of the latter terms come from the fact that there are multiple ways to remove a trace from a tableau with more than two columns. Recall that the irreducible $SO(1,d-1)$ representations are traceless. The decomposition of $U^{\threetwoone}$ can then be written
\begin{equation}
\begin{split}\label{eq:3,2,1_SO_decomp_tensors}
    U^{\threetwoone}_{\mu\nu\alpha|\beta\gamma|\delta} &= u^{\threetwoone}_{\mu\nu\alpha|\beta\gamma|\delta} + \bigg( \frac{1}{2} \eta_{\beta[\mu} u^{\twooneone}_{\nu\alpha]|\gamma|\delta} - (\beta\leftrightarrow\gamma) \bigg) + u^{\threeone}_{\mu\nu\alpha|[\beta} \eta_{\gamma]\delta} + \eta_{\delta[\mu} u^{\twotwo}_{\nu\alpha]|\beta\gamma} \\
    &\quad + \bigg( \frac{1}{2} \eta_{\beta[\mu|}\eta_{\gamma|\nu} u^{\oneone}_{\alpha]|\delta} - (\beta\leftrightarrow\gamma) \bigg) + \eta_{\delta[\beta}\eta_{\gamma][\mu} u^{\two}_{\nu\alpha]} + \bigg( \frac{1}{2} \eta_{\delta[\mu|} \eta_{\beta|\nu} u'^{\,\oneone}_{\alpha]|\gamma} - (\beta\leftrightarrow\gamma) \bigg) \\
    &\quad + u'^{\,\two}_{[\mu\nu} \eta_{\alpha][\beta}\eta_{\gamma]\delta} + \bigg( \frac{1}{2}\eta_{\delta[\mu|}\eta_{\beta|\nu} u''^{\,\oneone}_{\alpha]|\gamma} - (\beta\leftrightarrow\gamma) \bigg) + \bigg( \frac{1}{2} \eta_{\beta[\mu|}\eta_{\gamma|\nu} \eta_{\alpha]\delta} u^{\bullet} - (\beta\leftrightarrow\gamma) \bigg) \\
    &\quad +\bigg( \frac{1}{2}\eta_{\delta[\mu|}\eta_{\beta|\nu}\eta_{\alpha]\gamma} u'^{\,\bullet} - (\beta\leftrightarrow\gamma) \bigg) + \bigg( \frac{1}{2} \eta_{\delta[\mu|}\eta_{\beta|\nu}\eta_{\alpha]\gamma} u''^{\,\bullet} - (\beta\leftrightarrow\gamma) \bigg)
\end{split}
\end{equation}
where all the $u$ tensors are traceless. We will not give explicit formulae for the $u$ tensors, but they are constructed by Young projection of $U^{\threetwoone}$ onto the irreducible $SO(1,d-1)$ representations. In particular, each $u$ is traceless. Schematically, each $u$ tensor contains some number of traces of $U^{\threetwoone}$. This is a generalisation of the standard decomposition of a symmetric 2-tensor into irreducible $SO(1,d-1)$ tensors: $A_{\mu\nu} = (A_{\mu\nu} - \frac{1}{d}\eta_{\mu\nu}A\indices{_\sigma^\sigma}) + \frac{1}{d}\eta_{\mu\nu}A\indices{_\sigma^\sigma}$.

Note that there are three $u$ tensors in the $\oneone$ representation which we denote separately by $u^{\oneone}$, $u'^{\,\oneone}$ and $u''^{\,\oneone}$. 
This reflects the three ways in which this representation can arise when two traces are removed from the $\threetwoone$ representation. Indeed, this gives the multiplicity in eq.~\eqref{eq:3,2,1_SO}. 
Similarly, there are two terms in the $\two$ representation and three terms in the trivial $\bullet$ representation.

We first note that not all of the irreducible $SO(1,d-1)$ representations will contribute to $\Omega[V]$. Namely, the $\threeone$, $\oneone$, $\two$, and $\bullet$ representations vanish by $R_{\mu\nu}=0$ when contracted with $R^{\alpha\beta\gamma\delta}$ and so give $\Omega[V]=0$. Thus, the only $SO(1,d-1)$ representations which potentially contribute to non-trivial $\Omega[V]$ are $\threetwoone$, $\twooneone$ and $\twotwo$, so we restrict to
\begin{equation}\label{eq:3,2,1_SO_restricted}
    U^{\threetwoone}_{\mu\nu\alpha|\beta\gamma|\delta} = u^{\threetwoone}_{\mu\nu\alpha|\beta\gamma|\delta} + \bigg( \frac{1}{2} \eta_{\beta[\mu} u^{\twooneone}_{\nu\alpha]|\gamma|\delta} - (\beta\leftrightarrow\gamma) \bigg) + \eta_{\delta[\mu} u^{\twotwo}_{\nu\alpha]|\beta\gamma} 
\end{equation}
The goal of the decomposition into irreducible $SO(1,d-1)$ representations was to allow for the first term of eq.~\eqref{eq:divOmega_app} to vanish by the field equation $R_{\mu\nu}=0$. However, when we contract eq.~\eqref{eq:3,2,1_SO_restricted} with $\partial^\mu R^{\alpha\beta\gamma\delta}$, none of the terms vanish by using $R_{\mu\nu}=0$ or the Bianchi identities in eq.~\eqref{eq:bianchis}. Therefore, these representations cannot be allowed in $V$ if we want $\Omega[V]$ to be conserved. Therefore, the irreducible $GL(d,\mathbb{R})$ representation $\threetwoone$ does not give rise to any non-zero conserved 2-forms $\Omega[V]$. 

We now proceed to the $\twotwotwo$ $GL(d)$ representation in eq.~\eqref{eq:dRV_decomp}, which decomposes into irreducible $SO(1,d-1)$ representations as
\begin{equation}\label{eq:2,2,2_SO_decomp}
    \twotwotwo_{\,GL} = \left( \; \twotwotwo \oplus \twooneone \oplus\two \oplus \oneone \oplus \bullet \right)_{SO}
\end{equation}
Note that, while one may na\"{i}vely think that there are multiple traces which can be removed from the $\twotwotwo$ representation, they are not independent. This follows from the irreducibility of the $\twotwotwo$ representation, which requires that $W^{\twotwotwo}$ is invariant under permutations of the three columns. Therefore, it is equivalent to take a trace between any two columns of this representation. This is why there are no representations in the decomposition \eqref{eq:2,2,2_SO_decomp} with multiplicity greater than one.

One can then expand the $GL(d,\mathbb{R})$ tensor $W^{\twotwotwo}$ according to the decomposition \eqref{eq:2,2,2_SO_decomp}, in a similar manner to eq.~\eqref{eq:3,2,1_SO_decomp_tensors}. 
\begin{equation}
\begin{split}\label{eq:W_2,2,2_explicit}
    {W^{\twotwotwo}_{\mu\nu|\alpha\beta|}}^{\gamma\delta} &= {w^{\twotwotwo}_{\mu\nu|\alpha\beta|}}^{\gamma\delta} + {w^{\twooneone}_{\mu\nu|[\alpha|}}^{[\gamma} \delta_{\beta]}^{\delta]} + {w^{\twooneone}_{\alpha\beta|[\mu|}}^{[\gamma} \delta_{\nu]}^{\delta]} + \bigg(\frac{1}{2} {w^{\twooneone\,\gamma\delta}}_{\mu|[\alpha} \eta_{\beta]\nu} - (\mu\leftrightarrow\nu)\bigg) \\
    &\quad + w^{\two}_{\mu\nu} \delta_{[\alpha}^{[\gamma} \delta_{\beta]}^{\delta]} + w^{\two}_{\alpha\beta} \delta_{[\mu}^{[\gamma} \delta_{\nu]}^{\delta]} + \bigg(\frac{1}{2}w^{\two\gamma\delta} \eta_{\alpha[\mu} \eta_{\nu]\beta} - (\alpha\leftrightarrow\beta)\bigg) \\
    &\quad + {w^{\oneone}_{[\mu|}}^{[\gamma} \eta_{\nu][\alpha} \delta_{\beta]}^{\delta]} + \delta_{[\mu}^{[\gamma} w^{\oneone}_{\nu]|[\alpha} \delta_{\beta]}^{\delta]} + {w^{\oneone}_{[\alpha|}}^{[\gamma} \eta_{\beta][\mu} \delta_{\nu]}^{\delta]} + \bigg(\frac{1}{2}\eta_{\mu[\alpha} \delta_{\beta]}^{[\gamma} \delta^{\delta]}_\nu w^{\bullet} - (\mu\leftrightarrow\nu)\bigg)
\end{split}
\end{equation}
where all the $w$ tensors are traceless and are constructed by Young projection of $W^{\twotwotwo}$ onto the $SO(1,d-1)$ representations. We have raised the final two indices using $\eta^{\mu\nu}$ in order to write the anti-symmetrisations in a more compact form.
The different terms corresponding to the same Young tableau are necessary as the $\twotwotwo$ representation is symmetric under permutations of the three columns.

Again, we are only interested in those representations which contribute non-trivially to $\Omega[V]$. By contracting eq.~\eqref{eq:W_2,2,2_explicit} with $R^{\alpha\beta\gamma\delta}$, the $\oneone$ and $\bullet$ representations give a vanishing result from $R_{\mu\nu}=0$. Therefore, the $SO(1,d-1)$ representations which contribute to $\Omega[V]$ are $\twotwotwo$, $\twooneone$ and $\two$. 

We must determine which of these representations can yield a conserved 2-form $\Omega[V]$, which requires that the first term of eq.~\eqref{eq:divOmega_app} vanishes. Contracting eq.~\eqref{eq:W_2,2,2_explicit} with $\partial^\mu R^{\alpha\beta\gamma\delta}$, the $\twotwotwo$ and $\twooneone$ representations do not give a vanishing result. Therefore, they cannot appear in $V$.
Therefore, the only $SO(1,d-1)$ representation which gives a non-zero contribution to $\Omega[V]$ and also makes the first term in eq.~\eqref{eq:divOmega_app} vanish is $\two$. 

In summary, we have shown that any $V$ in the $\fourtwo_{GL}$ representation will give rise to a non-zero conserved 2-form $\Omega[V]$. The only other representation which can contribute to a non-zero conserved $\Omega[V]$ is the $SO(1,d-1)$ representation $\two$ which is contained in the decomposition of the $GL(d,\mathbb{R})$ representation $\twotwotwo$.

\paragraph{Representations giving independent $\Omega[V]$.}

While the $\two$ representation appearing in the decomposition \eqref{eq:W_2,2,2_explicit} can give rise to non-zero $\Omega[V]$, we now show that this $\Omega[V]$ can also be produced by a $\fourtwo$ tensor $V$. 
Contracting eq.~\eqref{eq:W_2,2,2_explicit} with the Riemann tensor, we find the 2-form $\Omega[V]$ corresponding to the $\two$ representation is
\begin{equation}
    \Omega[w^{\two}] = 2R_{\mu\nu\alpha\beta} w^{\two\,\alpha\beta}
\end{equation}
which is of the form of the Penrose 2-form in eq.~\eqref{eq:Penrose2form}.
However, we know from section~\ref{sec:Penrose_special_case} that the Penrose 2-form can also be produced by a $\fourtwo$ tensor $V$ given by eq.~\eqref{eq:V_Penrose}. 
It follows that the full set of possible conserved 2-forms $\Omega[V]$ can be accounted for solely by the $\fourtwo$ $GL(d)$ representation. 
As emphasised throughout, our intention is to classify the conserved 2-forms $\Omega[V]$, rather than the tensors $V$ themselves.
Therefore, for our purposes it is sufficient to restrict $V$ to transform in this representation. As mentioned at the beginning of this appendix, it remains to ensure that the second term in eq.~\eqref{eq:divOmegaWorking} vanishes in order for $\Omega[V]$ to be conserved. This puts a differential condition on $V$. This is considered in the main text, and the details can be found in Appendix~\ref{app:ansatz}.

\section{Constraint on $V$ from representation theory}
\label{app:ansatz}

In this appendix we give a derivation of the constraint given in eq.~\eqref{eq:final_V_ansatz} via representation theory arguments. We require that $\Omega[V]$ is conserved for some \pqi{4,2}-tensor $V$. From eq.~\eqref{eq:divOmegaWorking3}, this requires
\begin{equation}\label{eq:divOmega_appendix}
    0 = \partial^\mu \Omega[V]_{\mu\nu} = R^{\alpha\beta\gamma\delta} \partial^\mu V_{\mu\nu\alpha\beta|\gamma\delta}
\end{equation}
which imposes a differential constraint on $V$. We begin with a $GL(d,\mathbb{R})$ decomposition of the tensor product \pq{3,2} representation in which $\dd_L^\dag V$ transforms:
\begin{equation}\label{eq:ddag_V_decomp}
    \dd_L^\dag V \in \ydiagram{1,1,1} \otimes \ydiagram{1,1} = \ydiagram{1,1,1,1,1} \oplus \ydiagram{2,1,1,1} \oplus \ydiagram{2,2,1}
\end{equation}
However, using the fact that $V$ transforms in the \emph{irreducible} \pqi{4,2} $GL(d)$ representation, we have
\begin{equation}
    V_{[\mu\nu\alpha\beta|\gamma]\delta} = 0 \implies V_{\mu[\nu\alpha\beta|\gamma\delta]} = 0 \implies \partial^\mu V_{\mu[\nu\alpha\beta|\gamma\delta]}=0
\end{equation}
such that the $\five$ representation does not contribute to $\dd_L^\dag V$ in eq.~\eqref{eq:ddag_V_decomp}. The remaining contributions to $\dd_L^\dag V$ can further be decomposed into $SO(1,d-1)$ representations as
\begin{equation}\label{eq:divV_reps}
    \dd_L^\dag V \in \left( \; \ydiagram{2,1,1,1} \oplus \ydiagram{1,1,1} \; \right)_{SO} \oplus \left( \; \ydiagram{2,2,1} \oplus \ydiagram{2,1} \oplus \ydiagram{1} \; \right)_{SO} 
\end{equation}
where the $SO(1,d-1)$ representations are traceless with respect to the $SO(1,d-1)$-invariant metric $\eta_{\mu\nu}$.
We write this decomposition explicitly as
\begin{equation}
    \partial^\mu V_{\mu\nu\alpha\beta|\gamma\delta} = \frac{1}{2} \left( A^{\fourone}_{\nu\alpha\beta\gamma|\delta} + \eta_{\gamma[\nu} B^{\three}_{\alpha\beta]\delta} + C^{\threetwo}_{\nu\alpha\beta|\gamma\delta} + \eta_{\gamma[\nu} D^{\twoone}_{\alpha\beta]|\delta} + \eta_{\gamma[\nu}E^{\one}_\alpha \eta_{\beta]\delta} \right) - (\gamma\leftrightarrow\delta)
\end{equation}
where the tensors $A$, $B$, $C$, $D$ and $E$ are in the corresponding representations in eq.~\eqref{eq:divV_reps} (in particular, they are traceless).
Inserting this decomposition into eq.~\eqref{eq:divOmega_appendix} and using the field equation and Bianchi identity \eqref{eq:EOM_and_Bianchi}, we find that the $A$, $B$ and $E$ contributions vanish, leaving
\begin{equation}
    \partial^\mu \Omega[V]_{\mu\nu} = R^{\alpha\beta\gamma\delta} C^{\threetwo}_{\nu\alpha\beta|\gamma\delta} + \frac{1}{3} R\indices{^{\alpha\beta}_\nu^\delta} D^{\twoone}_{\alpha\beta|\delta}
\end{equation}
This vanishes for all field configurations $h_{\mu\nu}$ only if $C=D=0$. That is, these representations cannot appear in $\dd^\dag_L V$ if $\Omega[V]$ is to be conserved. 
Therefore, the allowed $SO(1,d-1)$ representations in the decomposition of $\dd_L^\dag V$ are
\begin{equation}
    \dd_L^\dag V \in \ydiagram{2,1,1,1} \oplus \ydiagram{1,1,1} \oplus \ydiagram{1}
\end{equation}
This then yields precisely the constraint which $V$ must satisfy in eq.~\eqref{eq:final_V_ansatz}.

\bibliographystyle{JHEP}
\bibliography{references}

\end{document}